\documentclass[12pt]{article}

\usepackage{amssymb}
\usepackage{amsfonts}
\usepackage{amsmath}
\usepackage{amsthm}

\title{Contextual Measurement Model and Quantum Theory}

\author{ Andrei Khrennikov\\ 
Linnaeus University, International Center for Mathematical Modeling\\  in Physics and Cognitive Sciences
 V\"axj\"o, SE-351 95, Sweden\\
email: Andrei.Khrennikov@lnu.se}

\begin{document}
\maketitle
\date{}

\abstract{We develop the contextual measurement model (CMM) which is used for clarification of the quantum foundations. This model matches with Bohr's views on the role of experimental contexts. CMM is based on contextual probability theory which is connected with generalized probability theory. CMM covers measurements in classical, quantum, and semi-classical physics. The CMM formalism is illustrated by a few examples. We consider CMM framing of classical probability, the von Neumann measurement theory, the quantum instrument theory.  CMM can also be applied outside of physics, in cognition, decision making, and psychology, so called quantum-like modeling.} 

{\bf keywords:} contextual measurement model, quantum foundations,  contextual probability, generalized probability,  von Neumann measurements, quantum instruments, quantum-like modeling

\section{Introduction}

Interrelation of quantum and classical probability theories is very complex foundational issue, involving interpretational, 
mathematical, and philosophic questions. Research in this area is characterized by the diversity of views, opinions, and mathematical formalisms (see, e.g., \cite{VN0}-\cite{Holik}). We remark that generally quantum mechanics (QM) is characterized by the diversity of interpretations.     

My own understanding is that quantum probability is a machinery for probability update, analogous to classical Bayesian inference \cite{KHR2}-\cite{HAVKHRQ1}. In contrast to the latter, quantum probability inference is not based on the Bayes formula for conditional probability. Quantum probability theory is a theory of probability inference with a special class of probability update transformations given by projections or quantum instruments. It is natural to create a general probabilistic framework that covers both the classical and quantum ones. Such generalization can come with up a global panorama as from the top of a mountain one can enjoy a panorama of the whole city and through this panorama connect districts which otherwise look as totally separated. In this way it is easier to fiend similarities and differences in districts plans and architecture of buildings. As just one of the possible machines for probability update, the quantum probabilistic formalism  would lose its mystery.

One of such ``panoramic frameworks'' is the {\it contextual measurement model} (CMM) based on contextual probability space. Its development was initiated in \cite{KHR2}, continued in a series of author's works (e.g., \cite{KHR3a}-\cite{KHR_interference}) and summarized in monograph \cite{KHR_CONT}. In these studies the main emphasis was to modification of {\it the formula of total probability} (FTP) -- its transformation into FTP with an interference term, expressing interference of probabilities, e.g., in the two slit experiment. In my previous studies contextual probability approach was partially shadowed by appeal to von Mises frequency theory of probability \cite{VM,INT} and realization of experimental contexts as von Mises collectives.  

In this paper CMM's development is continued to towards the abstract contextual formalization of other basic features of quantum probability as order and response replicability effects in sequential measurements, entanglement, 
the violation of the Bell inequalities and establishing coupling with quantum instruments theory as well as with {\it linear space representation} (LSR) of generalized probability theory.   

CMC is the basis of the {\it V\"axj\"o interpretation} of QM \cite{KHR2}, \cite{Vaxjo2002}-\cite{V2}, \cite{KHR_CONT} - one of the contextual probabilistic interpretations. Since probability update is at the same time information update, the V\"axj\"o interpretation is part of the information interpretation of QM. This paper presents CMM consistently in the most general form by highlighting its basic properties, interference of probabilities, order effect, entanglement, violation of the Bell inequalities.

The abstract CMM formalism is illustrated by a few examples. We start with CMM framing of classical probability theory 
(Kolmogorov \cite{K,KE}) serving as the basis of classical statistical physics and thermodynamics. Then we consider the 
von Neumann \cite{VN0,VN}  quantum measurement theory with observables given by Hermitian operators and the state update of the projective type and represent it as CMM. The quantum instrument theory is generalization of the von Neumann theory permitting state updates of the non-projective type and it is also can be represented as CMM.   We  also show connection of the generalized probability theory with the state space consisting of probability measures with CMM. Finally, LSR for contextual probability space is constructed by using the construction going back to Mackey.

CMM can also be applied outside of physics, in so called quantum-like modeling (see, e.g., books \cite{QL1,Open_KHR}). This the rapidly developing area of research stimulated by the recent quantum information revolution. In  quantum-like modeling the quantum methodology is applied  to cognition, decision making, psychology, game theory, economics and finance, and  AI. Universally, quantum-like models need not be based on the complex Hilbert space formalism. They can employ other contextual probability calculi and CMMs \cite{QL1}.

\subsection{Contextuality of probability}

 From the mathematical viewpoint, the essence of the problem is in generalizations of conditional probability and probability update. On the way to such rethinking of the interrelation between the classical and quantum probability theories, I was strongly influenced by Ballentine who treated all quantum probabilities as conditional probabilities \cite {Accardi2,Accardi3,BL,BL2,BL2a}. Later I learned that this was also  Koopman's viewpoint  \cite{Koopman}. It is interesting that  Kolmogorov (who in 1933 formalized classical probability in the measure-theoretic framework \cite{K,KE}) advertised this conditional viewpoint even for classical probability. This viewpoint was especially clearly described in his early works preceding  monograph \cite{K}. Unfortunately, these works (in Russian and published in proceedings)  
are practically unknown, see \cite{INT} for references and details. I got to know about these ``pre-axiomatic'' studies of Kolmogorov from Shiryaev and Bulinski,  former students of Kolmogorov. But even monograph  \cite{K} (see also \cite{KE}) contains the statement which in modern words can be formulated as a statement about {\it contextuality of probability.}  Kolmogorov's message is that it is meaningless to speak about probability without determining a complex of experimental conditions, measurement context. This Kolmogorov position matches well with Bohr's statement on contextuality of measurement's outcomes that is the cornerstone of his complementarity principle \cite{BR0} (which is better to call contextuality-complementarity principle \cite{NL0B,ABELL}).

Unfortunately, Kolmogorov's original message on contextuality of probability was practically ignored in further development of classical probability theory. A mathematical work on probability is  typically started with fixing one probability measure, without mentioning that it corresponds to some measurement context. The contextuality component of Bohr's statements on complementarity was neither emphasized in quantum foundational research; typically the Bohr complementarity principle is reduced to the wave-particle duality.      

Following Bohr \cite{BR0}, Kolmogorov \cite{K,KE}, Koopman \cite{Koopman}, Accardi \cite{Accardi2,Accardi3}, and Ballentine \cite{BL0}-\cite{BL2a}, I  introduced the contextual probability space and CMM based on it \cite{KHR_CONT}. The main idea behind this approach is operating solely with contexts and observables and to exclude  physical systems from consideration. 

 A {\it measurement context} consists of a pre-measurement context $C,$ an observable $A,$  and post-measurement context $C_{A=x}$ corresponding to the outcome $A=x,$ i.e., a triple $(C, A, C_{A=x}).$ Transformation $C \to C_{A=x}$ can be described as a map $T_A(x): {\cal C} \to {\cal C},$  where ${\cal C}$ is the set of pre-measurement contexts. Alike theory of quantum instruments \cite{}, we call the pair $I_A=(A, T_A)$ a contextual instrument. The latter is the basic mathematical component of measurement theory. It is meaningless to formulate it solely in terms of observables. The same observable $A$ can be a component of a variety of instruments describing different measurement procedures for $A.$ So, an observable is a theoretical quantity expressing some features of pre-measurement contexts.

In QM  one operates with the notion of ``state'', not ``context''. These notions are similar, but have some inetrpretational differences (see Appendix 1 for the discussion). 

We also mention Feynman's contextual analysis of the two slit experiment in the book \cite{FeynmanP,Feynman}. He presented the purely probabilistic picture this fundamental experiment of QM and expressed
the interference phenomenon as interference of probabilities. Mathematically he described this situation as the violation of additivity of probability, the classical formula is disturbed by an additional term, the interference term. In classical probability the combination of additivity and the Bayes formula  for conditional probability leads to FTP, the formula of total probability, playing the important role in probability inference. Following this line, Feynman's conclusion can be rewritten as the violation of classical FTP. The difference between classical and quantum probability models can be moved from the violation of additivity of probability to the violation of  the Bayes formula for conditional probability - quantum probability is additive, but conditioning is not Bayesian. The quantum FTP is a perturbation of classical FTP with an additional term, the interference term. The main distinguishing feature of Feynman's presentation of the two slit experiment and generally quantum interference is its contextual structure, he operates with three contexts $C_1, C_2, C_{12},$ the first slit is open and the second is closed, vice verse, and both slits are open. Quantum probabilistic specialty is expressed not via LSR of states and observables, but in the purely contextual probabilistic way. 

We make a remark on the notion of contextuality. In the modern quantum information literature the notion of contextuality is reduced to contextuality of the joint measurement of a few quantum observables. This sort of contextuality was considered by Bell in \cite{Bell2,Bell1} in his analysis of the violation of the Bell inequalities (although he did not used the term ``contextuality''). (It seems that this term was introduced in the book of Beltrametti and Cassinelli \cite{Beltrametti}.)

Feynman's contexuality \cite{FeynmanP,Feynman} is more general and in fact coincides with Bohr's contextuality \cite{BR0}. In my works including those on the V\"axj\"o interpretation, I followed Bohr and Feynman: context as a complex of all physical conditions involved in an experiment.     

\subsection{Linear space vs. contextual frameworks for probability}

Our foundational pathway is opposite to the pathway leading to the generalized probability theories \cite{Ozawa}, 
\cite{Mackey0}-\cite{Ozawa}, \cite{Beltrametti,Svozil} (see, e.g., \cite{Holik} for a review) which are directed to creation of general  LSR, linear space representation, of probability and measurement process. LRS also provides a panoramic view which covers both classical and quantum probabilities and observables. This is the linear panorama illuminating the place of the quantum probability and measurement formalism among other linear models. 

We remark that there are several different approaches to generalized probability theory and corresponding measurement theory, but all of them are either equivalent or only slightly different. 
One of them is  the Davies-Lewis \cite{Davies-Lewis} operational probability theory grounded on LRS  with the base norm spaces, a class of partially ordered linear spaces. The corresponding measurement theory is formulated within instrument theory \cite{Davies-Lewis, 
DV, Ozawa, Oz1}; in particular, observables are mathematically represented as positive operator valued measures (POVMs). They are widely used  in quantum information theory \cite{O1}-\cite{O3}, \cite{dariano1}- \cite{dariano3}.
(see also \cite{KHRfrontiers}-\cite{Open_KHR} for applications to cognition and decision making). Another approach is to start with an abstract definition of state space, a convex subset of a linear space. It goes back to Gudder's work \cite{Gudder} who constructed the operational representation of quantum states and observables starting with pre-convex structure. Under natural condition, this approach leads to a convex state space and LRS for the latter.  As was shown by Ozawa \cite{Ozawa}, these two formulations (Davies-Lewis  and Gudder) are actually equivalent. In section \ref{CLPR} we explore the Davies-Lewis approach for  operational LRS of measurement model with states given by classical probability measures. Then we express this model in the form of CMM.
   
The most close to CMM is the model in that one starts with all possible probabilities that can be generated in an experiment, conditioned on preparation and measurement procedures (see,  Mackey's book \cite{Mackey}. 
Then one proceeds to LSR.  This construction can be employed to construct LSR for CMM (section \ref{MackeyS}).  However, this is done just to show connection with previously developed theories. \footnote{Another approach to generation of the complex Hilbert space representation of CMM is developed in series of author's works, see, e.g., \cite{KHR_iterference,LF,KHR_CONT}. It is based on the contextual version of FTP, FTP with interference term.}

CMM development is important for quantum foundations. CMM demystifies quantum theory by reducing its probabilistic counterpart to the tool for probability update and inference (cf. with QBism \cite{Fuchs1} - \cite{Fuchs5}); CMM diminishes the role of pure states, in the complete agreement with the statistical (ensemble) interpretation of QM; in  CMM  quantum interference is just an additive  perturbation of classical FTP due to interplay between a few measurement contexts; the violation of the Bell inequality has the same origin; contextual entanglement is naturally coupled to classical  dependence of random variables. The latter demystifies entanglement. This is very important for resolution of the one century long debate on quantum nonlocality. See Appendix 2 for comparison of CMMs with and without LSRs.   

On the other hand, LSR is very convenient from the mathematical viewpoint (simply linear algebra) and operating within the LSR framework is useful in concrete mathematical calculations. However, the calculations should be completed by the critical analysis on connection of the mathematical LSR-constructions with physics. In quantum-like modeling a similar problem arises - the problem of matching between the output of the Hilbert space formalism and some psychological effects in decision making 
\cite{WB}-\cite{OJMP}.  

Our CMM can be considered as the most general probabilistic framework for measurement, in particular, the notion of contextual probability space is based on the first three axioms of Mackey's theory \cite{Mackey}. Then Mackey moves towards quantum logic by constraining the model with additional axioms. This path makes theory mathematically elegant, but at the same time more complex and the basic probabilistic components are blurred by additional mathematical constructions.    
 
\section{Contextual Measurement Model (CMM)}
\label{CMM} 

\subsection{Contextual probability space}    
		
{\bf Definition 1.}		{\it A contextual probability space is a triple
$\Sigma= ({\cal C}, {\cal O}, {\cal P}),$
where ${\cal C}$ and ${\cal O}$ are sets of pre-measurement contexts and observables and ${\cal P}$ is the space of the corresponding probability distributions.}

In physics pre-measurement contexts can be associated with preparation procedures.\footnote{In applications outside of physics, so called quantum-like modeling, not all pre-measurement contexts can be straightforwardly represented in the form of a preparation procedure; here we operate with mental, social, financial pre-measurement contexts. Within the statistical (ensemble) interpretation of QM, contexts are represented as ensembles of similarly prepared systems.} Each observable $A$ has its range of values $X_A;$ for simplicity, consider discrete observables, i.e., having finite ranges of values. The following considerations are straightforwardly extended to observables with arbitrary ranges of values.  
For an observable $A$ and pre-measurement context $C,$  denote the probability of an outcome $x \in X_A$ as $ P_C^A(x) \equiv P_C(A=x).$  By definition of a probability distribution
\begin{equation}
\label{L2}
P_C^A(x) \geq 0, \sum_{x\in X_A} P_C^A(x)=1.
\end{equation}
The range set $X_A$ can be endowed with the algebra of all its subsets ${\cal F}_A.$ We
set $P_C^A (G)=\sum_{x \in G} P_C^A(x), G \in {\cal F}_A.$ This is a probability measure on  ${\cal F}_A.$   
In the definition of $\Sigma,$ the symbol ${\cal P}$ denotes the collection of such probability 
measures 
$$
{\cal P}=\{P_C^A:  C \in {\cal C}, A \in {\cal O}\}
$$
(see Axiom 1 in Mackey's book \cite{Mackey}).
Elements of ${\cal P}$ are called contextual probabilities. These are the analogs of the conditional probabilities in the classical (Kolmogorov \cite{KE}) probability model. But we reserve the term ``conditional probability'' for a special class of contextual probabilities generated by context's updates. 

It is natural to assume (see Axiom 2 in Mackey's book \cite{Mackey})  that two observables having the same probability distribution for all contexts  should coincide, i.e., 
\begin{equation}
\label{L2m}
P_C^{A_1}= P_C^{A_2} \; \mbox{ for any }   C \in {\cal C} \Rightarrow A_1=A_2. 
\end{equation}
We also assume (see Axiom 2 in Mackey's book \cite{Mackey}) that two contexts having the same probability distribution for all observables should coincide, i.e., 
\begin{equation}
\label{L2m1}
P_{C_1}^{A}= P_{C_2}^{A} \; \mbox{ for any }   A \in {\cal O} \Rightarrow C_1=C_2. 
\end{equation}

The average of an observable $A \in {\cal O}$ (with $X_A \subset \mathbb{R})$  w.r.t. a pre-measurement context $C \in {\cal C}$ is defined as
\begin{equation}
\label{L2aa}
\langle A\rangle_C\equiv E[A|C]= \sum_{x\in X_A} x P_C(A=x).
\end{equation}

\subsection{Context update and conditional probability}

Measurement of an observable $A$ with the concrete outcome $x$ in a pre-measurement context $C$ updates this pre-measurement context:
\begin{equation}
\label{L3s}
 C \to C_{A=x}.
\end{equation}
In terms of preparation procedures, we can consider a measurement as a subsequent preparation procedure; context
 $ C_{A=x}$ is measurement of $A$ and filtering w.r.t. the fixed outcome $x.$\footnote{In terms of ensembles, $A$-measurement is performed for systems of initially prepared ensemble and then systems generating the outcome $A=x$ form new ensemble. }
It is natural to consider this map only for contexts belonging to the set
\begin{equation}
\label{w}
{\cal C}_A(x)= \{C: P_C(A= x) > 0\}.
\end{equation}
If $P_C(A= x)=0,$ then the post-measurement context is not well defined.
Thus, each observable $A$ and its outcome  $x$ determine a map 
\begin{equation}
\label{L3}
T_A(x): {\cal C} \to {\cal C}, \; C \to C_{A=x}=T_A(x) C,
\end{equation}
with the domain of definition ${\cal C}_A(x).$  

The delicate point of measurement theory is that generally an observable does not determine the context update map unequally. An observable $A$ can be measured via different measurement procedures and each procedure generates its own context update map. A pair $I_A$= (observable, context update map)=($A, T_A)$ is called a {\it contextual instrument} (cf. section \ref{instrument});  a pair $(C, I_A),$ or a triple $(C, A, T_A),$ is called a {\it measurement context.}  We stress once again that a variety of instruments can be associated with the same observable $A: \;  I_A= (A, T_A), I_A^\prime= (A, T_A^\prime), I_A^{\prime\prime}= (A, T_A^{\prime\prime}),.... \;.$ 
We stress that all these update maps have the same domain of definition determined by the observer $A,$ see (\ref{w}).

Typically one fixes some class of context update maps. In the von Neumann \cite{VN0,VN}
measurement theory (section \ref{projection}), this is the class of normalized projections.   
In quantum instrument theory (section \ref{instrument}) these are quantum channels or more generally in the theory of Davis-Levis instruments, these are positive trace preserving maps.

We emphasize that von Neumann's measurement theory is very special: here by fixing an observable, a Hermitian operator $\hat A,$ we automatically fix the update map -- via operator's spectral family. This special situation leads to the illusion that an observable determines the update map. We repeat that generally this is not the case.     

\medskip

{\bf Definition 2.}		{\it Let $\Sigma = ({\cal C}, {\cal O}, {\cal P})$ be a contextual probability space. A contextual  measurement model (CMM) is a pair $M=(\Sigma, {\cal I}),$ where $\Sigma$  is a contextual probability space and ${\cal I}$ is a collection of contextual instruments.}

\medskip

CMM is a set of measurement contexts, i.e., triples $(C, A, T_A).$ CMM generates the notion of the conditional probability:

{\bf Definition 3.} {\it Consider a measurement context $(C, A, T_A).$ Let it generate the output $A=x$ and the corresponding context update, $C \to C_{A=x}=T_A(x) C.$ Consider measurement of another observable $B$ under condition $A=x,$ i.e., w.r.t. context $C_{A=x}.$ The conditional  probability is given by the formula:}
\begin{equation}
\label{L5}
P_{C,I_A}(B=y| A=x) \equiv P_{C_{A=x}}(B=y) =P_{T_A(x) C}(B=y), \; C \in {\cal C}_A(x).
\end{equation}

We note that this definition involves context update only for the $A$-observable; different contextual 
instruments $I_A, I^\prime_A,...,$ induce their own probabilistic conditioning, 
$P_{C,I_A}(B=y| A=x), P_{C,I^\prime_A}(B=y| A=x),... \;.$ For simplicity, we shall typically omit the index
 $I_A$ of dependence on the concrete instrument. 

\subsection{Contextual formula of total probability}
\label{CFTP}

Now we point out that generally contextual probability differs from the classical Kolmogorov probability \cite{KE}. One of the basic classical laws of probability is the law of total probability formulated in the form of FTP, {\it formula of total probability},
\begin{equation}
\label{L6}
P(B=y)=\sum_{x \in X_A}  P(B=y| A=x) P(A=x).
\end{equation}
In classical probability theory contextual probability is identified with conditional one (section \ref{KKK}) and the  
contextual-conditional analog of FTP has the form, 
\begin{equation}
\label{L6aa}
P_C(B=y)=\sum_{x \in X_A}  P_C(B=y| A=x) P_C(A=x).
\end{equation}
see (\ref{L7ss}). However, in a general contextual probability space this formula can be violated,  
\begin{equation}
\label{L7}
P_C(B=y) \not= \sum_{x \in X_A}  P_C(B=y|A=x) P_C(A=x) = 
\end{equation}
$$
\sum_{x \in X_A}  P_{C_{A=x}}(B=y) P_C(A=x).
$$
The difference between the LHS and RHS determines the degree of context-disturbance due to its update; it can serve as a measure of nonclassicality of a contextual model,
\begin{equation}
\label{L7aa}
\delta_C(B=y|A) = P_C(B=y) - \sum_{x \in X_A}  P_C(B=y| A=x) P_C(A=x).
\end{equation}
We call this quantity {\it the interference term} \cite{KHR2,KHR_CONT}. This consideration can be formalized in the contextual formula of total probability (FTP with an interference term):
\begin{equation}
\label{nm}
P_C(B=y)=\sum_{x \in X_A}  P_C(B=y| A=x) P_C(A=x) + \delta_C(B=y|A). 
\end{equation}
The equality of the interference term to zero is the necessary condition of the classical probabilistic representation of a contextual probability model, but it is not the sufficient condition \cite{KHR_CONT}.

Let us jump for the moment to section \ref{projection} in that the von Neumann quantum measurement model is treated as CMM.
Consider dichotomous observables $A=x_1,x_2$ and $B=y_1, y_2.$  In this case the interference term has the form
$$
\delta_C(B=y|A) =  
$$
\begin{equation}
\label{nm2}
2 \cos \theta \sqrt{ P_C(B=y| A=x_1) P_C(A=x_1) P_C(B=y| A=x_2) P_C(A=x_2)},
\end{equation}
where context $C$ is identified with quantum state $\psi$ (for simplicity consider contexts corresponding to pure states) and 
the angle $\theta=\theta(B=y|A; \psi).$ For dichotomous observables, even in general CMM it is useful to write the interference term as 
$$
\delta_C(B=y|A) = 
$$
\begin{equation}
\label{nm3}
 2 \lambda \sqrt{ P_C(B=y| A=x_1) P_C(A=x_1) P_C(B=y| A=x_2) P_C(A=x_2)},
\end{equation}
where $\lambda=\lambda(B=y|A;C).$ If $|\lambda| \leq 1,$ then this is the trigonometric interference and the interference term 
has the form (\ref{nm2}). If  $|\lambda| \geq 1,$ then this is the hyperbolic interference  and the interference term can be represented as 
$$
\delta_C(B=y|A) =
$$
\begin{equation}
\label{nm4}
  \pm 2 \cosh \sqrt{P_C(B=y| A=x_1) P_C(A=x_1) P_C(B=y| A=x_2) P_C(A=x_2)}.
\end{equation}
In the quantum framework such interference can be generated by quantum instruments 
 \cite{LF}. In general CMM we can employ the hyperbolic version of QM \cite{KHR_CONT}.

\subsection{Conditional JPD and order effect}

For observables $A_1, A_2 \in {\cal O},$ the conditional joint probability distribution (JPD) is defined by 
\begin{equation}
\label{L11}
 P_C(A_1=x_1,A_2=x_2) = P_C(A_1=x_1)P_C(A_2=x_2| A_1=x_1).
\end{equation}
We remark that this is really a probability distribution, i.e., 
$\sum_{x_1,x_2} P_C(A_1=x_1,A_2=x_2) =1.$ We can also define JPD for inverse order of measurements,  
$P_C(A_2 = x_2, A_1=x_1) = P_C(A_2=x_2)P_C(A_1=x_1| A_2=x_2).$ Observables $A_1,A_2$ show the {\it order effect} (OE) w.r.t. context $C,$ if 
\begin{equation}
\label{L12}
 P_C(A_1=x_1,A_2=x_2) \not= P_C(A_2 = x_2, A_1=x_1),
\end{equation}
for at least one pair of outcomes $(x_1, x_2);$ otherwise,  there is no OE in context $C.$

We remark OE was actively investigated in decision making and psychology, both theoretically and experimentally; in particular, within  quantum-like modeling -- the applications of quantum methodology and formalism to decision making and psychology  \cite{WB}-\cite{OJMP}.  

\subsection{Conditional Compatibility} 

In the absence of OE we have:
\begin{equation}
\label{L12aa}
 P_C(A_1=x_1,A_2=x_2) = P_C(A_2 = x_2, A_1=x_1), \; x_j \in X_{A_j},
\end{equation}
i.e., 
\begin{equation}
\label{b}
P_C(A_1=x_1) P_C(A_2 = x_2| A_1=x_1)= P_C(A_2=x_2) P_C(A_1 = x_1| A_2=x_2).
\end{equation}
In this case we call the observables {\it conditionally compatible} for context $C\in {\cal C}$ and their JPD is defined by (\ref{L12aa}). We remark that the marginals of JPD coincide with the probability distributions 
 $P_C^{A_i}.$ 

In the von Neumann CMM $M_{\rm{QVN}}$ (section \ref{projection}) with observables and state updates of the projection type, conditional  compatibility for all possible pre-measurement contexts (given by density operators) is equivalent to 
commonly considered compatibility of observables and their  representation by commutative Hermitian  operators. 

By considering conditional JPD,   we do not assume that the observables $A_1$ and $A_2$ are jointly measurable. We consider sequential measurements, say first $A_1$ then $A_2$ or vice verse. We remark that, in fact, precisely this experimental setup is realized in the Bell experiments. Here the instances of time for  measurement's outputs  on subsystems coincide with zero probability; always the click of the photo-detector for the subsytem $S_1$ is before the click of the photo-detector for the subsystem $S_2$ or vice verse and the time window serves for clicks pairing.
   
Conditional compatibility implies the Bayes formula for the conditional probability:
\begin{equation}
\label{b1}
P_C(A_2 = x_2| A_1=x_1)= \frac{P_C(A_1=x_1,A_2=x_2)}{P_C(A_1=x_1)}, 
\end{equation}
\begin{equation}
\label{b1t}
P_C(A_1 = x_1| A_2=x_2)= \frac{P_C(A_1=x_1,A_2=x_2)}{P_C(A_2=x_2)}.
\end{equation}

The equality (\ref{b}) implies the Bayes theorem for the probability inference. Let
the outcomes of the observable $A_2$ label some hypotheses, $H_1,..., H_{m}.$
Then (\ref{b}) is written as
\begin{equation}
\label{b3}
P_C(A_2= H_j| A_1=x_1)= \frac{P_C(A_2= H_j) P_C(A_1 = x_1| A_2= H_j)}{P_C(A_1=x_1)}.
\end{equation} 
The Bayes formula for conditional probability (\ref{b1}),(\ref{b1t})  implies the validity of the classical FTP,
i.e., the Bayes theorem can be written in the standard form: 
\begin{equation}
\label{b4}
P_C(A_2= H_j| A_1=x_1)= \frac{P_C(A_2=H_j) P_C(A_1 = x_1| A_2= H_j)}
{\sum_{H_i} P_C(A_1=x_1|H_i) P_C(A_2=H_i)}.
\end{equation} 

\subsection{Replicability and response replicability}

Observable $A$ shows {\it replicability} for context $C,$ 
if 
\begin{equation}
\label{L8}
P_C(A=x, A=x) = P_C(A=x),  
\end{equation}
or 
\begin{equation}
\label{L8m}
P_{C_{A=x}}(A=x) =1.
\end{equation}
Observable $A$ shows replicability if (\ref{L8}) holds for any $C\in {\cal C}_A(x), x \in X_A.$  

In quantum-like modeling, the following effect plays the important role.
Observables $A_1$ and $A_2$ show {\it the response replicability effect} (RRE) w.r.t. context $C,$ if 
\begin{equation}
\label{L8k}
P_C(A_1=x_1, A_2=x_2, A_1=x_1) = P_C(A_1=x_1, A_2=x_2),
 \end{equation}
\begin{equation}
\label{L8n}
P_C(A_2=x_2, A_1=x_1, A_2=x_2) = P_C(A_2=x_2, A_1=x_1)
 \end{equation}
for all pairs of outcomes $(x_1, x_2).$
This is a kind of the memory effect. The challenging problem for quantum-like modeling was the combination of OE and RRE \cite{PLOS}. It was solved in articles \cite{ENTROPY,OJMP} within quantum instrument theory.

\subsection{Correlations and Bell type inequalities}

Consider a pair of conditionally compatible  observables $A,B \in {\cal O}$ (with $X_A, X_B \subset \mathbb{R}).$ Their correlation   w.r.t. a  context $C \in {\cal C}$ is defined as
\begin{equation}
\label{L2bb}
\langle A B\rangle_C\equiv E[AB|C]= \sum_{x,y} x y P_C(A=x,B=y).
\end{equation}

The most popular Bell type inequality is the CHSH inequality. We consider this inequality within CMM. 
There are given four observables  $A_i, B_j, i,j=1,2,$ valued in $[-1,1];$ observables in the pairs $(A_i,B_j)$ are conditionally compatible for some context $C$ with JPDs  $P_C^{A_i,B_j}.$ The CHSH inequality has the form:  
\begin{equation}
\label{L2cc}
|\langle A_1 B_1\rangle_C + \langle A_2 B_1\rangle_C + \langle A_1 B_2\rangle_C - \langle A_2 B_2\rangle_C| \leq 2, 
\end{equation}
i.e., 
\begin{equation}
\label{L2ccg}
|\sum_{x,y} x y (P_C^{A_1,B_1}(x,y) + P_C^{A_2,B_1}(x,y) + P_C^{A_1,B_2}(x,y) - P_C^{A_2,B_2}(x,y))|\leq 2.
\end{equation}
If there exists a probability measure $P_C$ such that JPDs $P_C^{A_i,B_j}$ can be obtained as its marginals, e.g.,
$$
P_C^{A_1,B_1}(x,y)= \sum_{x_2,y_2} P_C(x,x_2, y, y_2),
$$
then the CHSH inequality holds true. However, if such $P_C$ does not exist, then this inequality can be violated and the maximum of its left-hand side w.r.t. contexts $C \in {\cal C}$ and observables $A_1,A_2, B_1,B_2 \in {\cal O}$ valued in $[-1,1]$ can approach the value  four, $\max_{\rm{CHSH}}=4.$ This maximum depends on CMM. For von Neumann CMM $M_{\rm{QVN}},$ this is $\max_{\rm{CHSH}}=2\sqrt{2}.$ 

\subsection{Functions of observables}

Suppose that all observables are valued in multidimensional real space and we remove (for the moment) the restriction that observables' ranges of values are finite. We consider probability measures on the 
$\sigma$-algebra ${\cal B}$ of the Borel sets; it is generated by all semi-open intervals,
$(\alpha_1, \beta_1] \times ... \times (\alpha_n, \beta_n].$ So,
$P_C^A$ is a probability measure on ${\cal B}.$
A function $f: \mathbb{R}^n \to  \mathbb{R}^m$ is 
called ${\cal B}$-measurable if for any Borel subset $G$ of $\mathbb{R}^m$ its preimage 
$f^{-1}(G)$ is a Borel subset of $\mathbb{R}^n.$ Only such functions are considered. 

 Following Mackey \cite{Mackey} (Axiom 3), we assume that for each  $A\in {\cal O}$ with the range of values $\mathbb{R}^n$ and a Borel function $f: \mathbb{R}^n \to  \mathbb{R}^m$ there exists $B=B_f \in {\cal O}$ such that, for any $C \in {\cal C}$ and Borel set $G \subset \mathbb{R}^m,$ 
\begin{equation}
\label{ttq}
P_C^B(G)=P_C^A(f^{-1}(G)).
\end{equation} 
Such observable is uniquely defined, due to condition (\ref{L2m}) (Mackey's Axiom 2) and it can be denoted 
as $B=f(A).$

Two observables $A_1$ and $A_2$ are called functionally compatible (jointly measurable) if there exists an observable 
$A$ and functions $f_i$ such that $A_i= f_i(A).$ For the von Neumann CMM $M_{\rm{QVN}}$ (section \ref{projection}) functional 
compatibility is equivalent to compatibility and hence conditional compatibility. Generally in CMM interrelation between these two notions is complex and  we shall not proceed to detailed comparison.

\subsection{Entanglement of contextual instruments}
\label{ENTAB}

Entanglement is typically considered as one of the distinguished features of LSR; from my viewpoint the association of entanglement with LSR and the tensor product structures shadows its physical nature; its mathematical description is identified with physics. As was shown in articles \cite{ENT,ENT1}, the notion of entanglement can be formalized in the purely probabilistic framework and dissociated from the tensor product and generally from LSR.  

By starting with such probabilistic approach to the notion of entanglement, the authors of \cite{ENT,ENT1} proceed towards its complex Hilbert space realization. Now we present the purely probabilistic picture of entanglement. The main value of the contextual probabilistic realization of entanglement is in clarification of its foundational meaning.
At the same time, the use of LSR can essentially simplify the concrete calculations. However, one should be careful with connection of the mathematical structures of LSR with physics (or in quantum-like modeling 
with e.g. psychology and decision making).

Consider CMM  $M=(\Sigma, {\cal I}),$ where $\Sigma=  ({\cal C}, {\cal O}, {\cal P})$  is a contextual probability space and ${\cal I}$ is a collection of contextual instruments of this model, i.e., pairs 
(observable, state update map). Consider two contextual instruments $I_A=(A, T_A)$ 
and $I_B= (B, T_B).$

{\bf Definition 4.}    {\it In pre-measurement context $C\in {\cal C},$ the outcome $B=\beta$ depends on the outcomes of $A$ if for at least two values of $A, \alpha= \alpha_i, \alpha_j,$
the corresponding conditional probabilities don't coincide,} 
\begin{equation}
\label{G1}
P_C( B= \beta| A= \alpha_i) \not= P_C( B= \beta| A= \alpha_j).
\end{equation} 

Thus,  the probability to get the outcome  $B=\beta$ if  the preceding $A$-measurement had the outcome $A=\alpha_i$ differs from the probability to get the same outcome $B=\beta$ if  the preceding $A$-measurement had the outcome $A=\alpha_j.$ We remark that the update map $T_B$ is not involved in this definition, i.e., one can consider just an observable $B$ without referring  to the corresponding instrument $I_B.$ We consider two instruments by symmetry reason. 

We note that the outcome $B=\beta$ does not depend on the outcomes of the observable $A$ iff
\begin{equation}
\label{G1q}
P_C( B= \beta| A= \alpha_i) = P_C( B= \beta| A= \alpha_j),\; \mbox{for all pairs}  \;\alpha_i, \alpha_j,
\end{equation} 
i.e. the conditional probability for this outcome is constant w.r.t. the outcomes of $A.$ 
Denote it $P_C( B= \beta| A).$ The following natural question arises:  Does the probability $P_C( B= \beta| A)$ coincide with unconditional probability $P_C(B= \beta)?$

{\bf Definition 5.}    {\it Two instruments $I_A$ and $I_B$  are  called $AB$-entangled in $C \in {\cal C}$ or $C$ is $AB$-entangled,  if all outcomes of the $B$-observable depend on outcomes of $A$-observable, i.e. for all $\beta$ condition (\ref{G1}) holds  for some  $ \alpha_i, \alpha_j.$}

Concerning the notation ``$AB$-entangled'', it would be better to write ``$I_AI_B$-entangled'', to emphasize 
that this is entanglement of instruments and not simply the observables, but to make the notation compact, we proceed with ``$AB$-entangled''. The order of observables is important. Generally $AB$-entanglement does not imply $BA$-entanglement. This is the purely  probabilistic definition, it does not involve LSR, it can be applied to any statistical physical theory. This definition formalizes dependence of observables. We introduce the following quantitative measure of entanglement: 

{\bf Definition 6.} {\it   For contextual instruments $I_A$ and $I_B$ and pre-measurement context $C,$   
$AB$-concurrence of conditional probabilities is defined as}
\begin{equation}
\label{ME}
\lambda_{AB}(\psi)= \sum_\beta\sum_{\alpha \not=\alpha^\prime} 
|P_C( B= \beta| A= \alpha) - P_C(B= \beta| A= \alpha^\prime)|.
\end{equation} 
 
The crucial issue is that $AB$-concurrence depends on a pair of instruments.

{\bf Proposition 1.} {\it For dichotomous  observables $A, B= \pm,$  dependencies  of
 the values  $B=-$ and $B=+$  on the outcomes of $A$ are equivalent. Thus, each dependence is equivalent to $AB$-entanglement.}   
\medskip

{\bf Proof.} In the state context $C,$ the value $B=-$ depends on the outcomes of $A$ if 
\begin{equation}
\label{G3}
P_C( B= -| A= +) \not= P_C( B=  - | A= -),
\end{equation} 
This automatically implies that even the value $B=+$ depends on the outcomes of $A,$
$$
P_C( B= +| A= +) = 1- P_C( B= -| A= +)   \not=  
$$
$$
1- P_C( B= -| A= -)  = P_C( B=  + | A= -),
$$
i.e., instruments $I_A$ and $I_B$ are entangled in context $C.$ 

\medskip

As was already pointed out, in articles \cite{ENT,ENT1}  contextual probabilistic entanglement can be 
realized in the complex Hilbert space and in this way connected with the ordinary notion of entanglement. In the LSR representation, the main distinguishing feature of $AB$-entanglement (Definition 5) is that it is associated with a pair of instruments, $I_A, I_B.$ The standard definition of entanglement is coupled with the tensor product structure and not with two concrete 
instruments (observables).  

For simplicity consider CMM $M_{\rm{QVN}}$ (section \ref{projection})  with von Neuman observables \cite{VN}; in this CMM an observable $\hat A,$ a Hermitian operator, automatically determines the state update map, via its spectral family. So, there is no need to operate with instruments, one can operate solely with observables. In this CMM (which is typically used in entanglement studies), contextual probabilistic entanglement is associated with the pairs of observables, i.e., two observables are entangled or disentangled in some state (context) $C=\hat \rho.$ The main mathematical features of $AB$-entanglement and ordinary, tensor product based, entanglement are similar, but some essential differences can be found \cite{ENT,ENT1}.  

The probabilistic viewpoint on the ``EPR-paradox'' \cite{EPR} is presented in  Schr\"odinger's  papers  \cite{SCHE,SCHEa} that initiated the modern theory of entanglement. However, this theory ignores the important message of Schr\"odinger: entanglement characterizes probability update for the outcomes  of observable $B$ conditioned on the  outcomes of observable $A.$ In the framework of \cite{SCHE,SCHEa}, it is meaningless to speak about entanglement  without specifying the observables. The state update -- the Hilbert space representation of the probability update -- encodes the procedure of conditional prediction. For Schr\"odinger, the quantum formalism is a mathematical machinery for probability prediction (as in the 
V\"axj\"o interpretation or QBism) and a quantum state is a part  of such machinery. We can say that Schr\"odinger interpreted quantum probabilities  as conditional (contextual) probabilities and entanglement as contextual probabilistic entanglement. But this is my private interpretation of  Schr\"odinger's views and many experts in quantum foundations may disagree with me.  

By following Schr\"odinger \cite{SCHE,SCHEa}, in article \cite{ENT} we considered a special sort  of the contextual probabilistic entanglement that matches perfectly the Schr\"odinger analysis of the EPR-argument.  

{\bf Definition 7.}  {\it For $C\in {\cal C},$  instruments  $I_A$ and $I_B$ 
are perfectly  conditionally correlated for values  $(A=\alpha, B=\beta)$ if the conditional probability to get the outcome  $B=\beta$ if  the preceding $A$-measurement had the outcome $A=\alpha$ equals to 1,}
\begin{equation}
\label{Lx3}
P_C(B=  \beta| A= \alpha) =  1.
\end{equation}

More generally consider observables with values $(\alpha_i)$ and $(\beta_i)$ and some set $\Gamma$ of pairs $(\alpha_i, \beta_j).$   

{\bf Definition 7a.}  (EPR-entanglement) {\it Let $C\in {\cal C}.$  If instruments $I_A$ and $I_B$ are perfectly  conditionally correlated for all pairs belonging to $\Gamma,$
then they are called EPR-entangled w.r.t. set $G$ in context $C.$}    

 We are interested in sets $\Gamma$ such that each of $\alpha$ and $\beta$ values appears in the pairs once and only once.  We call such  EPR-entanglement complete.

For example, for two dichotomous observables with $\alpha, \beta=\pm,$ 
we consider, e.g.,  the set of the  pairs $(A=+, B=-), (A=-, B=+),$ in short, $A=-B$ EPR-entanglement, 
or the  pairs $(A=+, B=+), (A=-, B=-),$  $A=B$ EPR-entanglement. We analyze such EPR-entanglements. 

Let us start with $A=-B$ EPR-entanglement, i.e., 
$P_C(B=-|A=+)=1$ and $P_C(B=+|A=-)=1.$ Thus, $P_C(B=+|A=+)=0$ and $P_C(B=-|A=-)=0,$ and
$P_C(B=-|A=+)=1 \not= P_C(B=-|A=-)=0$ and    $P_C(B=+|A=-)=1 \not= P_C(B=+|A=+)=0.$ In this case EPR-entangled instruments are automatically entangled in the sense of Definition 5.

Now turn to  $A=B$ EPR-entanglement, i.e., 
$P_C(B=+|A=+)=1$ and $P_C(B=-|A=-)=1.$ Thus, $P_C(B=-|A=+)=0$ and $P_C(B=+|A=-)=0,$ and
$P_C(B=+|A=+)=1 \not= P_C(B=+|A=-)=0$ and    $P_C(B=-|A=-)=1 \not= P_C(B= -|A=+)=0.$ And again EPR-entangled instruments are automatically entangled in the sense of Definition 5.

Thus, EPR-entanglement is just the very special case of $AB$-entanglement.  

For dichotomous observables, $AB$-concurrence of conditional probabilities (\ref{ME}) has the form 
$
{\cal Con}_{AB}(\psi)= |P( B=+| A= -) - P( B= +| A= +)| +  |P( B=-| A= -) - P( B= -| A= +)|,
$
hence it can written  as 
\begin{equation}
\label{G4ee}
\lambda_{AB}(\psi) = 2|P( B=+| A= -) - P( B= +| A= +)|. 
\end{equation}
From this formula, we immediately obtain the following characterization of maximally $AB$-entangled states: 

{\bf Proposition 4.} {\it $AB$-concurrence of conditional probabilities approaches its maximal value, 
$\lambda_{AB}(\psi)=2,$ if and only if the instruments are EPR-entangled in pre-measurement context $C.$} 
  
\subsection{Distinguishing features of contextual measurement models}

We list the probabilistic constraints which can be used to distinguish different CMMs (theoretically and experimentally):
\begin{itemize}
\item violation of FTP
\item  OE
\item  RRE
\item OE+RRE 
\item violation of Bell inequalities
\end{itemize}

\subsection{Interpretations of contextual probability}

Probability is characterized by the diversity of interpretations \cite{INT}. We now discuss the 
interpretations of contextual probability. We start with the remark that mathematically a contextual probability space $\Sigma = ({\cal C}, {\cal O}, {\cal P})$ cannot be described as the single Kolmogorov probability space: $\Sigma$ is a bunch of such spaces
However, fixed   $C\in {\cal C}$  and  $A \in {\cal O}$ can be realized within some probability space $K=(\Omega, {\cal F}, P)$ with realization of observable $A$ by a random variable $a: \Omega \to X_A;$  its  
probability distribution coincides with $P_C^A,$ i.e., 
$$
P_C^A(\alpha)= P(\omega\in \Omega: a(\omega)=\alpha), \; \alpha \in X_A.
$$ 
We note that  a contextual probability space $\Sigma$ can be represented as 
\begin{equation}
\label{jj}
 \Sigma= \cup_{C,A} K_{C,A},
\end{equation}
where $K_{C,A}$ is a Kolmogorov probability space for describing measurement of the
observable $A$ in the pre-measurement context $C.$

Therefore one can assign to  the contextual probability any interpretation used for probability defined in the measure-theoretic framework. The main interpretation employed in classical and quantum physics is the frequency interpretation. In the Kolmogorov theory \cite{K,KE}, this interpretation is mathematically rooted to the strong law of large numbers.  Another basic interpretation of probability in physics is the statistical (ensemble) interpretation. By this interpretation $\Omega= \Omega_C$ represents 
an ensemble of systems prepared for measurement  and the probability measure $P=P_A$ depends on the 
observable $A.$   Finally, we mention the subjective interpretation. It is widely employed in decision making and psychology, but was not used in physics until QBism was invented.         

In contextual probability,  we need not represent a pre-measurement context $C$  by an ensemble of systems. 
Instead, we can consider a sequence of measurements of an observable $A$ in the same pre-measurement context $C,$ 
\begin{equation}
\label{VM1}
{\bf x}\equiv {\bf x}_{C,A}= (x_1,..., x_N,...,...),
\end{equation}
where $x_j (\in X_A=\{\alpha_1,...,\alpha_m\})$ are measurement's outcomes. Such sequence determines the frequencies of realizations of the concrete values, 
\begin{equation}
\label{VM2}
\nu_N (\alpha_j)=  n_N(\alpha_j)/N,
\end{equation}
where $\nu_N (\alpha_j)$ is the number of measurement's outcomes with the fixed value $\alpha_j.$ The probability to obtain the value $\alpha_j$ in a sequence of  measurements ${\bf x}$ is defined as the limit
\begin{equation}
\label{VM13}
P_C^A(\alpha_j)\equiv \lim_{N \to \infty} \nu_N (\alpha_j).
\end{equation}
This is the straightforward frequency introduction of probability. The deep mathematical theory of frequency probability was developed by von Mises \cite{VM} (see also \cite{INT} for an introduction). A sequence ${\bf x}$ generated in by observations is called a {\it collective.} Von Mises' theory is a theory of collectives.
Instead of operations on sets, as done in the Kolmogorov measure-theoretic theory, von Mises constructed a probability theory based on operations with collectives. We remark that 
$P_C(A=\alpha_j)\equiv P_C^A(\alpha_j)$ can be considered as the probability generated by the 
collective ${\bf x}\equiv {\bf x}_{C,A},$ i.e.,   $P_C^A(\alpha_j)= P_{{\bf x}_{C,A}}(\alpha_j).$ It is important to note that not all collectives are compatible - combinable in von Mises' terminology. Two collectives 
${\bf x}$ and ${\bf y},$  are combinable if their  combination   
\begin{equation}
\label{VM4}
{\bf z} =(z_1,...,z_N,...), \; z_j=(x_j,y_j),
\end{equation}
is also a collective  and the probability distributions  $P_{{\bf x}}$ and $P_{{\bf y}}$ are marginal for the probability distribution $P_{{\bf z}},$ i.e.,
\begin{equation}
\label{VM5}
P_{{\bf x}}(\alpha_j)= \sum_{\beta_i} P_{{\bf z}} (\alpha_j,\beta_i), \; 
P_{{\bf z}}(\beta_i)= \sum_{\alpha_j} P_{{\bf z}} (\alpha_j,\beta_i). 
\end{equation} 

The frequency probability theory contains the notion of conditional probability that is similar to CMM's
conditioning, the post-measurement context $C_{A=a}$ corresponds to the post-measurement collective 
${\bf x}_{C,A=a}$ (see \cite{VM,INT}  for the details).

In contrast to the Kolmogorov measure-theoretic probability theory, in the von Mises frequency probability theory classical FTP is violated \cite{INT,KHR_CONT}, the probabilities can interfere and generate the aditive perturbation of FTP in the form of the interference term. The presence of incombinable collectives leads to the violation of the Bell type inequalities \cite{INT,KHR_CONT}. 

The notion of collective was the seed for future growing theory of random sequences. Besides the existence of the limits for frequencies, (\ref{VM13}), a collective is characterized by stability of these limits w.r.t. place selections within a sequence  ${\bf x}$, i.e., the limit-probability is the same for all subsequences of ${\bf x}$ for a special class of place selections. However, von Mises' definition of 
place selection was criticized for non-rigorousness. Its critical analysis was very fruitful and led to the modern theory of randomness, see,e.g., monograph \cite{KHRWS}). In particular, the monograph presents 
a ``light version'' of the von Mises theory. In physics (at least in quantum physics) one does not analyze  in the random structure of the sequences of measurement's outcomes. In principle, in QM one can proceed with ``light-collectives'' determined solely by the existence of limits  (\ref{VM13}). The calculus of 
such ``light-collectives'' can be explored for the description of the probabilistic structure of QM \cite{KHRWS}. The first version of the contextual probability was presented in such ``light frequency'' framework.
    
Finally, we note that von Neumann \cite{VN0,VN} pointed to the von Mises \cite{VM} frequency probability as the probabilistic foundation for QM. This is the complex foundational issue.

\section{Contextual Measurement Model for Kolmogorov Theory}
\label{KKK}

Let $K= (\Omega, {\cal F}, P)$ be a Kolmogorov probability space \cite{KE}. Here $\Omega$ is a set of any origin, ${\cal F}$ is a collection its subsets forming $\sigma$-algebra, i.e., ${\cal F}$ is closed w.r.t. countable unions and intersections, and the operation of complement. (If $\Omega$ is finite, then ${\cal F}$ is collection of all its subsets.)  And $P$ is a probability measure on ${\cal F}.$

Set ${\cal C}=\{C \in  {\cal F}: P(C) \not=0\}.$ This is the set of contexts.
For each context $C,$ the Bayes formula defines the conditional probability measure 
$$
P_C(G)=P(G\cap C)/P(C), G \in {\cal F}.
$$

We highlight that the statistical mixtures of contexts are not determined, i.e., for subsets $C_1, C_2$  
of $\Omega$ and  weights $p_1, p_2 \geq 0, p_1+p_2=1,$ there is no a subset $C$ of $\Omega$  which can be identified with the weighted sum $p_1 C_1+p_2 C_2.$ 

As an illustrative example, consider some agricultural  region $\Omega$ and as contexts consider its sub-fields (some areas). Generally there is no field of the form $p_1 C_1+p_2 C_2.$  In applications to decision making and cognition,  one can meet the situations such that $p_1 C_1+p_2 C_2$ is determined only for a few  pairs of weights $p_1, p_2.$ This situation is related to poorness of the set of possible experimental contexts. 

The set of observables ${\cal O}$ is the set of (discrete) random variables, $a: \Omega \to X_a,$ where $X_a$ is a finite set. (Discrete random variables are considered for simplicity.) Denote this set by the symbol 
${\cal R}_d.$  For $x \in X_a,$ we set $\Omega_{a=x}= \{\omega \in \Omega: a(\omega)=x \}.$
Contextual probability coincides with conditional probability given by the Bayes formula:
\begin{equation}
\label{K1}
P_C^a(x)\equiv P_C(a=x)= P(\Omega_{a=x}\cap C)/P(C).
\end{equation}
Thus, the set of probability distributions ${\cal P}=\{P_C^a: C \in {\cal C}, a \in {\cal O} \}.$

For any set $D \in {\cal F}$ and random variable $a \in {\cal R}_d, x \in X_a,$ we define set 
$$
D_{a=x}= \{\omega \in D: a(\omega)=x \} = D \cap \Omega_{a=x}
$$ 
and the map
\begin{equation}
\label{L3pa}
T_a(x): {\cal F} \to {\cal F}, \; D \to D_{a=x}=T_a(x) D.
\end{equation}
For any context $C \in {\cal C}$ and random variable $a \in {\cal R}_d, x \in X_a,$ we define the family of contexts  
$$
{\cal C}_a(x)= \{C \in {\cal C}: P_C(a = x) > 0\}= \{C \in {\cal C}: P(C_{a=x}) > 0\}.
$$
Each random variable $a$ and its outcome  $x$ determine a map 
\begin{equation}
\label{L3p}
T_a(x): {\cal C} \to {\cal C}, \; C \to C_{a=x}=T_a(x) C,
\end{equation}
with the domain of definition ${\cal C}_a(x).$ 

Thus, classical CMM $M_{\rm{cl}}$ consists of measurement contexts composed of pre-measurement contexts - elements of ${\cal F}$ with non-zero probabilities, observables - (discrete) random variables, and  context update maps, ${\cal T}=\{ T_a(x) \}.$ Here each observable $a,$ random variable, determines uniquely the context update maps $T_a(x),$ and, hence, the contextual instrument.   

The conditional  probability is given by the Bayes formula:
$$
P_C(b=y| a=x)= P(\Omega_{b=y}\cap  \Omega_{a=x} \cap C)/P(\Omega_{a=x} \cap C)
$$
$$
= P(\Omega_{b=y}\cap  C_{a=x})/P(C_{a=x})=P_{C_{a=x}} (b=y).
$$

Since, for each $C \in {\cal C},$ $P_C$ is a probability measure, for any pair of random variables $a,b$, we have the following version of FTP, formula of total probability, section \ref{CFTP},
\begin{equation}
\label{L7ss}
P_C(b=y) = \sum_{x \in X_a}  P_C(b=y|a=x) P_C(a=x) = 
\end{equation}
$$
\sum_{x \in X_a}  P_{C_{a=x}}(b=y) P_C(a=x).
$$ 

In this measurement model all observables are compatible, conditional JPD coincides with JPD (again by Bayes formula); no pair of observables show OE, since 
$$
P_C(a_1=x_1, a_2=x_2)=P_C(a_2=x_2, a_1=x_1)=
$$
$$
P_C(\omega \in \Omega: a_1(\omega) =x_1, a_2(\omega)=x_2).
$$
All observables show repeatability, since $P_C(a=x, a=x)= P_C(a=x),$ and RRE, since  
$$
P_C(a_1=x_1, a_2=x_2, a_1=x_1)=P_C(\omega \in \Omega: a_1(\omega) =x_1, a_2(\omega)=x_2, a_1(\omega) =x_1) =
$$
$$
P_C(a_1=x_1, a_2=x_2).
$$
The Bell inequalities are not violated, since their derivation is based on the existence of JPD.

We can summarize properties of classical CMM  by referring to the aforementioned  list  of possibilities:
\begin{itemize}
\item violation of FTP - no
\item  OE - no
\item  violation of replicability - no
\item  RRE - yes
\item OE+RRE - yes 
\item violation of Bell inequalities - no
\end{itemize}

One of the problems of the above contextual representation of the classical probability is that the uniqueness conditions (\ref{L2m}), (\ref{L2m1}), Mackey's axiom 2, can be violated, i.e., generally  
\begin{equation}
\label{L2ma}
P_C^{a_1}= P_C^{a_2} \; \mbox{ for any }   C \in {\cal C} \not\Rightarrow a_1=a_2. 
\end{equation}
\begin{equation}
\label{L2m1a}
P_{C_1}^{a}= P_{C_2}^{a} \; \mbox{ for any }   a \in {\cal O} \not\Rightarrow C_1=C_2. 
\end{equation}
This problem can be easily resolved in the standard way, see below. 

Let us consider a Kolmogorov probability space $K= (\Omega, {\cal F}, P)$ with a complete probability measure, i.e., any subset of a set $D \in {\cal F}, P(D)=0,$ also belongs to   ${\cal F}.$  We recall that 
the symmetric difference of two sets $D_1$ and $D_2$ is defined as 
$$
D_1 \Delta D_2 = (D_1 \setminus D_2) \cup (D_2 \setminus D_1) = (D_1 \cup D_2) \setminus (D_1 \cap D_2).
$$ 
For $D_1,D_2 \in {\cal F},$  we set $D_1 \sim D_2$ if $P(D_1 \Delta D_2)=0.$ 
This is an equivalence relation on ${\cal F};$ it splits ${\cal F}$ into disjoint equivalence classes. 
Denote the set of equivalence classes by the symbol $\tilde{{\cal F}};$ denote the equivalence class of 
zero probability sets by the symbol $\tilde{Z}.$ The set of pre-measurement contexts is  $\tilde{{\cal C}}= \tilde{{\cal F}} \setminus \tilde{Z},$  i.e., all equivalent classes of sets from ${\cal F}$ of non-zero measure. 

We also modify the class of observables. Two random variables are equivalent, $a_1 \sim a_2,$ if 
$P(\omega \in \Omega: a_1(\omega) \not=  a_2(\omega))=0.$ This is the equivalence relation on the space of random variables, in our consideration these are discrete random variables ${\cal R}_d.$ So, ${\cal R}_d$ is split into disjoint classes of equivalent random variables, denote the set of these classes  by the symbol 
$\tilde{{\cal R}}_d$ and set $\tilde{{\cal O}}= \tilde{{\cal R}}_d.$ 

For a pre-measurement context $\tilde{C} \in \tilde{\cal C}$ and observable $\tilde{a} \in \tilde{\cal R}_d$ we define the probability distribution $P_{\tilde{C}}^{\tilde{a}}(x)= P_C(a=x)$ for some representatives $C \in  \tilde{\cal C}$
and $a \in \tilde{\cal R}_d$ (the correctness of this definition is proved below), and set $\tilde{\cal P}= \{ P_{\tilde{C}}^{\tilde{a}} \}.$
The modified classical contextual probability space is the triple  $\tilde{\Sigma}=(\tilde{\cal C}, \tilde{\cal R}_d, \tilde{\cal P}).$  

The map $T_a(x),$ see (\ref{L3p}), generates a map of  $\tilde{{\cal F}}$ 
into itself
\begin{equation}
\label{L3pqq}
\tilde{T}_a(x): \tilde{{\cal F}} \to \tilde{{\cal F}}.
\end{equation}
Set $\tilde{\cal C}_a(x)=\{ \tilde C\in \tilde {\cal C}:  \tilde{T}_a(x) \tilde C \in \tilde{\cal C} \}.$ Then $\tilde{T}_a(x): \tilde{\cal C}_a(x) \to \tilde{\cal C}.$  A measurement context is a triple 
$(\tilde{C}, \tilde{a}, \tilde{T}_a).$  
Modified classical CMM $\tilde M_{\rm{cl}}$ is given by the set of such measurement contexts.  

Now we demonstrate that in $\tilde M_{\rm{cl}}$ the uniqueness conditions (\ref{L2m}), (\ref{L2m1}),
Mackey's axiom 2,  hold true. We assume that the ranges of values of random variables are subsets of a set $X;$ for simplicity let $X=\mathbb{R}.$ 
First we note that if $D_1, D_2 \in \tilde{D},$ then $P(D_1)=P(D_2).$ We have
$P(D_1)= P((D_1 \cap D_2) \cup (D_1 \setminus D_2))=  P(D_1 \cap D_2)= P(D_2).$ Let now $C_1, C_2 \in \tilde{C},$ then, as we have seen  $P(C_1)=P(C_2);$ also, for any $G \in {\cal F}, P(G \cap C_1)= P(G \cap C_1\cap C_2)= P(G \cap C_2).$ Hence, $P_{C_1}(G)=P_{C_2}(G).$ So, conditional probability measure $P_C$ does not depend on the choice of a representative $C \in \tilde{C}$ and it can denoted as  $P_{\tilde{C}}.$

We show that implication (\ref{L2m}) holds. Let for random variables $a_1, a_2,$
$P_C^{a_1}= P_C^{a_2}$ for any context $C.$ Let, for some $x, P(\Omega_{a_1=x}) >0.$ Take
$C=\Omega_{a_1=x},$ then $$ 1=P_{\Omega_{a_1=x}}(\Omega_{a_1=x})= P(\Omega_{a_2=x} \cap \Omega_{a_1=x})/P(\Omega_{a_1=x}),$$ i.e.,  $$
P(\Omega_{a_2=x} \cap \Omega_{a_1=x})= P(\Omega_{a_1=x})=   
P((\Omega_{a_2=x} \cap \Omega_{a_1=x})\cup (\Omega_{a_1=x} \setminus \Omega_{a_2=x}).
$$ 
Hence,
$P(\Omega_{a_1=x} \setminus \Omega_{a_2=x})=0$ and symmetrically  
$P(\Omega_{a_2=x} \setminus \Omega_{a_1=x})=0.$ The sets $\Omega_{a_i=x}, i=1,2,$ belong to the same equivalence class. This implies that the random variables also belong to the same equivalence class.      

Now we turn  to implication (\ref{L2m1}). Let, for any random variable $a,$ 
$P_{C_1}^a= P_{C_2}^a.$ Select $a$ as the characteristic function of the set $C_1.$ Then $P_{C_1}^a(1)= 1=P(C_1\cap C_2)/P(C_2),$ i.e., 
$P(C_1\cap C_2) =P(C_2)=P((C_1\cap C_2) \cup (C_2\setminus C_1),$ i.e., $P(C_2\setminus C_1)=0$ and symmetrically  $P(C_1\setminus C_2)=0,$ i.e., contexts belong to the same equivalence class.

\section{Contextual Measurement Model for von Neumann Observables}
\label{projection}

We restrict consideration to finite dimensional Hilbert state spaces. 
The space of pre-measurement contexts is mathematically represented as the space of density operators ${\cal D}$, i.e., 
${\cal C}= {\cal D},$ observables are Hermitian operators (von Neumann observables \cite{VN0,VN}).  Denote the space of Hermitian operators by the symbol ${\cal L}_H,$ i.e., ${\cal O}= {\cal L}_H.$ This real linear space is endowed the with scalar product $\langle \hat A| \hat B \rangle = \rm{Tr} \hat A \hat B.$   

Operator $\hat A \in {\cal L}_H$ has the spectral decomposition: 
$\hat A =\sum_{x \in X_A} x\;  \hat E_A(x),$ where $\hat E_A(x)$ is the orthogonal projection on the subspace ${\cal H}_A(x)$ composed of eigenvectors with  eigenvalue $x,$ and $X_A$ is operator's spectral set.  Then
\begin{equation}
\label{L1q}
P_\rho^A(x) \equiv P_\rho(A=x) = \rm{Tr} \hat E_A(x) \hat \rho= \rm{Tr} \hat E_A(x) \hat \rho  \hat E_A(x),
\end{equation}
\begin{equation}
\label{L2q}
T_A(x): {\cal D} \to {\cal D}, \hat \rho_{A=x} \equiv T_A(x)\hat \rho=  \frac{\hat E_A(x) \hat \rho \hat E_A(x)}{\rm{Tr} \hat E_A(x) \hat \rho \hat E_A(x) },
\end{equation}
with the domain of definition ${\cal C}_A(x)=\{\hat \rho \in D: 
P_\rho(A =x) >0 \}.$ 

Thus, ${\cal P}=\{P_\rho^A: \rho \in {\cal D}, A \in {\cal L}_H \}$ and ${\cal T}=\{T_A(x): A \in {\cal L}_H, x \in X_A \}.$ We remark that observable $A$ uniquely determines  the family of maps $T_A(x), x \in X_A$ by equality (\ref{L2q}). So, measurement contexts can be represented by pairs  $(\hat \rho, \hat A), \hat \rho \in   {\cal D}, \hat A \in   {\cal L}_H.$ Denote this CMM by the symbol $M_{\rm{QVN}}.$

In this CMM one need not define separately context update maps, they are automatically encoded in  observables. On the one hand, this simplifies theory. On the other hand, this is the misleading path in measurement theory, cf. with quantum instrument theory.

 In  CMM $M_{\rm{QVN}},$ the conditional probability is given by the formula:
\begin{equation}
\label{L2qy}
 P_\rho(B=y|A=x)= \rm{Tr} \hat E_B(y) T_A(x) \rho=
\frac{\rm{Tr} \hat E_B(y)  \hat E_A(x) \hat \rho \hat E_A(x)}{\rm{Tr} \hat E_A(x) \hat \rho \hat E_A(x) }.
\end{equation}
It can be rewritten as 
\begin{equation}
\label{L2qymm}
 P_\rho(B=y|A=x)= \frac{\rm{Tr} \hat E_B(y)  \hat E_A(x) \hat \rho \hat E_A(x)\hat E_B(y)}{\rm{Tr} \hat E_A(x) \hat \rho \hat E_A(x)}.
\end{equation}
In this LSR-based CMM, it is convenient to introduce the maps:
\begin{equation}
\label{L2u}
{\cal I}_A(x) \hat \rho= \hat E_A(x) \hat \rho  \hat E_A(x)
\end{equation} 
Then the above formulas can be rewritten as
\begin{equation}
\label{L1v}
P_\rho(A=x) = \rm{Tr}  {\cal I}_A(x) \hat \rho,
\end{equation}
\begin{equation}
\label{L2j}
T_A(x)\hat \rho=  \frac{1}{P_\rho(A=x)} {\cal I}_A(x) \hat \rho.
\end{equation}
or 
\begin{equation}
\label{L2jj}
{\cal I}_A(x) \hat \rho=  P_\rho(A=x) T_A(x) \hat \rho;
\end{equation}
and the conditional probability is written in the form 
\begin{equation}
\label{L2qyb}
 P_\rho(B=y|A=x)= \frac{\rm{Tr} {\cal I}_B(y) {\cal I}_A(x) \hat \rho}{\rm{Tr}  {\cal I}_A(x) \hat \rho},
\end{equation}
and conditional JPD as
\begin{equation}
\label{L2qybhh}
 P_\rho(A=x, B=y)=  \rm{Tr} {\cal I}_B(y) {\cal I}_A(x) \hat \rho.
\end{equation}
These formulas lead to quantum instrument theory (section \ref{instrument}): $(A, {\cal I}_A(x))$ is a special quantum instrument  and $(A, T_A(x))$ is the corresponding contextual instrument. 

We note that in CMM $M_{\rm{QVN}}$ probabilities determine contexts (states) and observables (operators), i.e., (\ref{L2m}) and (\ref{L2m1}) hold (Mackey's axiom 2).

Let $P_{\rho}^{A_1}(x)= P_{\rho}^{A_1}(x)$ for all $\hat \rho \in {\cal D}$ and real $x.$
Then $\rm{Tr} (E_{A_1}(x)- E_{A_2}(x)) \hat \rho =\langle E_{A_1}(x)- E_{A_2}(x)| \hat \rho\rangle =0.$ 
Hence, $E_{A_1}(x)=E_{A_2}(x)$ for any $x;$ so $\hat A_1= \hat A_2.$ Thus, an observable can be identified with the set of probability distributions ${\cal P}_A=\{P_\rho^A: \rho \in {\cal D}\}.$

Now let  $P_{\rho_1}^{A}(x)= P_{\rho_2}^{A}(x)$ for all $\hat A \in {\cal L}_H$ and real $x.$ Then $\rm{Tr} E_{A}(x) (\hat \rho_1 - \hat \rho_2)= \langle E_{A}(x)| 
\hat \rho_1 - \hat \rho_2\rangle =0.$ Hence, $\hat \rho_1 =\hat \rho_2.$ Thus, a quantum state can be identified with the set of probability distributions ${\cal P}_\rho=\{P_\rho^A: A\in {\cal L}_H \}.$ 

In the von Neumann measurement theory, two observables are {\it compatible} if they are represented by commuting operators $\hat A, \hat B: \; [\hat A,\hat B]=0.$ Compatibility is interpreted as guarantying the possibility of joint measurement of these observables; their JPD is given by the formula: 
\begin{equation}
\label{JPD1}
P_\rho^{A, B} = \rm{Tr} \hat E_A(x) \hat E_B(y) \rho = \rm{Tr} \hat E_B(y)  \hat E_A(x) \rho. 
\end{equation}
In fact, this is the separate axiom - a complement to the Born rule \cite{VN}.  For compatible observables, JPD and conditional JPD coincide. The conditional JPD is given by
\begin{equation}
\label{JPD2}
P_\rho(A=x, B=y)= P_\rho(A=x) P_\rho(B=y|A=x)= 
\end{equation}
$$
P_\rho(A=x) P_{\rho_{A=x}}(B=y)=  
= \rm{Tr} \hat E_B(y)  \hat E_A(x) \rho  \hat E_A(x) =   
$$
$$
\rm{Tr} \hat E_A(x) \hat E_B(y)  \hat E_A(x) \rho=
\rm{Tr} \hat E_B(y)  \hat E_A(x) \rho.
$$

In particular, for compatible observables there is no OE for any state $\rho.$ If  operators $\hat A_1, \hat A_2,$ do not commute, then there exists a state $\hat \rho$ showing  OE for these observables, i.e., $P_\rho(A=x, B=y) \not= P_\rho(B=y,A=x).$ 

Each observable shows replicability, e.g.,  
$$
P_\rho(A=x, A=x)= \rm{Tr} \hat E_A(x) \hat E_A(x) \hat \rho \hat E_A(x)=   \rm{Tr} \hat E_A(x) \hat \rho= P_\rho(A=x).
$$ 
If observables are compatible, then for any $\hat \rho$ they show RRE, e.g.,  
$$
P_\rho(A_1=x_1, A_2=x_2, A_1=x_1) = \rm{Tr} \hat E_{A_1}(x_1) \hat E_{A_2}(x_2) \hat E_{A_1}(x_1) \hat \rho 
\hat E_{A_1}(x) \hat E_{A_2}(x_2)=
$$
$$
\rm{Tr} \hat E_{A_2}(x_2) \hat E_{A_1}(x_1) \hat \rho \hat E_{A_1}(x) =
P_\rho(A_1=x_1, A_2=x_2).
$$
We highlight that it is impossible to combine OE and RRE within  CMM $M_{\rm{QVN}}$
\cite{PLOS}. 

FTP (the formula of total probability) can be violated; classical FTP is additively perturbed by  the interference term; consider a pure state $|\psi\rangle,$ then 
\begin{equation}
\label{a7}
P_\psi(B= \beta)= \sum_{\alpha, \alpha^\prime} \langle \psi| \hat E_A(\alpha) \hat E_B(\beta) \hat E_A(\alpha^\prime) | \psi \rangle
\end{equation}
$$
= \sum_\alpha P_\psi(B= \beta|A= \alpha)P_\psi(A= \alpha) + \delta_\psi(B=y|A),
$$
where 
\begin{equation}
\label{aa7}
\delta_\psi(B=y|A)=\sum_{\alpha \not=\alpha^\prime} \langle \psi| \hat E_A(\alpha) \hat E_B(\beta) \hat E_A(\alpha^\prime) | \psi \rangle.
\end{equation}
In RHS, the first summand corresponds to classical FTP and the second one is the interference term; it quantifies the degree of non-classicality for this CMM. See section \ref{CFTP} for FTP in the general contextual probabilistic framework.

Consider dichotomous observables, $A=x_1,x_2$ and $B=y_1, y_2,$  of the von Neumann type. 
In this case the interference term has the form
$$
\delta_\psi(B=y|A) =  
$$
\begin{equation}
\label{nm2v}
2 \cos \theta \sqrt{ P_\psi(B=y| A=x_1) P_\psi(A=x_1) P_\psi(B=y| A=x_2) P_\psi(A=x_2)},
\end{equation}
where the angle $\theta=\theta(B=y|A; \psi).$

We can summarize properties of von Neumann CMM with the list  of possibilities presented above:
\begin{itemize}
\item violation of FTP - yes   
\item  OE - yes
\item  violation of replicability - yes
\item  RRE - no
\item  OE+RRE - no
\item violation of Bell inequalities - yes
\end{itemize}

\section{Contextual Measurement Model for Quantum Instruments} 
\label{instrument}

In this section we present CMM $M_{\rm{QI}}$ in that a measurement process is mathematically described by  quantum instrument theory
\cite{Davies-Lewis,DV}, \cite{Ozawa,Oz1}, \cite{O1}-\cite{dariano3}. This CMM extends CMM $M_{\rm{QVN}}$ in that a measurement process is mathematically described by a Hermitian operator (von Neumann observable).

The space of linear Hermitian operators ${\cal L}_H$ is a real Hilbert space. 
We consider linear operators acting in it to be {\it superoperators}. A superoperator $T$ is called positive if it maps the set of  positive semidefinite operators onto itself: for $g \geq 0, \; T(g) \geq 0.$

Consider an observable $A$ with finite range of  values $X.$ 
Its measurements can be performed with various apparatuses; for each apparatus, the corresponding measurement procedure is mathematically described in the following way.  

Any map $x \to {\cal I}(x),$ where for each $x \in X,$ the map  $ {\cal I}(x)$ is a positive superoperator and 
\begin{equation}
\label{sum}
{\cal I}(X)\equiv \sum_{x \in X} {\cal I}(x): {\cal D} \to {\cal D}
\end{equation}
is called a {\it quantum instrument.} It determines some observable; denote it by the symbol $A,$  
see mathematical formula (\ref{BRULEy3}).

 The probability of the output $A=x$ is given by the generalized Born rule in the form 
 \begin{equation}
\label{BRULEy}
P_\rho(A=x) = \rm{Tr}\; [{\cal I}(x) \rho].
\end{equation}

We note that measurement with the output $A=x$ generates the state-update by  transformation 
\begin{equation}
\label{TRA4}
\rho \to \rho_{A=x}= T_A(x) \rho \equiv \frac{{\cal I}(x)\rho}{Tr {\cal I}(x)\rho},
\end{equation}
with the domain of definition ${\cal C}_A(x)=\{\hat \rho \in D: 
P_\rho(A =x) >0 \}.$ 

Let 
\begin{equation}
\label{TRA4s}
{\cal I}(x)\rho= \hat E(x) \rho \hat E(x),
\end{equation}
where $(\hat E(x))$ are projections giving the orthogonal decomposition of $I.$ 
Such an instrument is called a projection instrument.    

The most natural generalization of projection instruments is an atomic instrument. Let $(\hat V(x))$ be a family of  linear operators contrained by the normalization condition:
\begin{equation}
\label{OPN2}  
\sum_x \hat  V(x) \hat V^\star(x) = I.
\end{equation}
An atomic quantum instrument is a super-operator of the form: 
\begin{equation}
\label{TRApr}
\rho \to {\cal I}(x)\rho= \hat V(x)\rho \hat V^\star(x). 
\end{equation}
Applications of the quantum instrument theory to quantum information are typically restricted by the use of 
atomic instruments.  

The space of Hermitian operators ${\cal L}_H$  is the real Hilbert space, i.e., for  each linear operator acting in ${\cal L}_H$ (superoperator) its adjoint is well defined. The 
generalized Born rule can written as  
\begin{equation}
\label{BRULEy1}
P_\rho(A=x) = \langle {\cal I}(x) \rho| I\rangle = \langle  \rho| {\cal I}^\star(x) I\rangle= 
\langle {\cal I}^\star(x) I| \rho \rangle, 
\end{equation}
where $I$ is the unit operator and ${\cal I}^\star(x)$ is superoperator that is adjoint to ${\cal I}(x)$ in Hilbert space ${\cal L}_H.$ Hence, the generalized Born rule has the form:
\begin{equation}
\label{BRULEy2}
P_\rho(A =x) = \rm{Tr} \hat A(x)  \rho, 
\end{equation}
where 
\begin{equation}
\label{BRULEy3}
\hat A(x) = {\cal I}^\star(x) I.  
\end{equation}
Operators $\hat A(x), x \in X,$ are called {\it effects}; they are positive semi-definite Hermitian and sum up to the unit operator:
$$
\sum_{x \in X} \hat A(x)=I.
$$ 
The family of operators $A =(\hat A(x), x\in X)$ is called a {\it positive operator valued measure} (POVM): for 
a subset $\Delta$ of $X,$ we set 
$$
\hat A(\Delta)= \sum_{x \in \Delta} \hat A(x) \geq 0;
$$ 
this is an additive operator-valued measure, i.e., for $\Delta_1, \Delta_2 \subset X, \Delta_1 \cap \Delta_2 = \emptyset,$ 
$$
\hat A(\Delta_1 \cup \Delta_2) = \hat A(\Delta_1) +  \hat A(\Delta_2).
$$ 
Instruments of the projection type, see (\ref{TRA4s}), determine the special class of POVMs, {\it 
projection valued measures} (PVMs).

Two POVMs $A =(\hat A(x), x \in X)$ and $B =(\hat B(y), y \in Y)$ are called {\it compatible} if there exists a POVM $C=(\hat C(x,y), (x,y) \in X \times Y)$ such that
\begin{equation}
\label{xy}
\hat A(x)= \sum_{y \in Y}  \hat C(x,y), \; \hat B(x)= \sum_{x \in X}  \hat C(x,y).
\end{equation}
Compatibility is interpreted as guarantying  the possibility of the joint measurement of these observables, their JPD is given by generalization of the Born rule for compatible von Neumann observables:
\begin{equation}
\label{xy1}
P_\rho^{A,B}(x,y)= \rm{Tr} \hat C(x,y) \hat \rho.
\end{equation}

In  contextual probability space, contexts are mathematically represented by density operators (quantum states),  ${\cal C}= {\cal D},$ observables by POVMs (also known as generalized quantum observables), the probability distributions are determined by the generalized Born rule (\ref{BRULEy}). CMM $M_{\rm{QI}}$ is endowed by quantum instrument maps updating  quantum states (contexts) due to the measurement feedback (\ref{TRA4}).

Consider POVM $\hat A =(\hat A(x))$ and all quantum instruments generating it via (\ref{BRULEy3}). Then the corresponding state (context) update  maps are defined by equality (\ref{TRA4}). The same POVM, generalized observable, can be coupled to a variety of such maps.  Therefore the commonly used approach highlighting POVMs as generalized observables is ambiguous. POVMs are just biproducts generated by quantum instruments.

Quantum instruments considered above were invented in article  \cite{Davies-Lewis} (see also monograph \cite{DV}) 
and these are {\it Davies-Levis instruments}, so $M_{\rm{QI}}= M_{\rm{QI; DL}}.$ In quantum information theory, one uses the special class of quantum instruments given by {\it completely positive maps} ${\cal I}(x);$ denote the corresponding  CMM 
$M_{\rm{QI;O}},$ where I use the index ``O'' to mention Masanao Ozawa who contributed so much into theory of 
such quantum instruments \cite{Ozawa,Oz1}, \cite{O1}-\cite{O3}. It is commonly assumed that the instruments belonging to
$M_{\rm{QI; DL}} \setminus M_{\rm{QI;O}}$ are non-physical. I debated this question with Masanao Ozawa and he firmly stays on this position. As was proved by him only completely positive instruments 
can be realized via the indirect measurement scheme \cite{Ozawa}. This scheme is adequate to 
quantum measurement processes and any deviation from this scheme is non-physical. Nevertheless, it might be that in quantum-like modeling, even instruments which are not completely positive can find applications. Such applications would lead to modification of the indirect measurement scheme, amy be via consideration of non-unitary interactions.   

We remark that instrument maps ${\cal I}(x)$ are linear in the Hilbert space ${\cal L}_H.$ In terms of context (state) update these maps they can be written as
\begin{equation}
\label{SC}
{\cal I}(x)= P_\rho(A=x) T_A(x).
\end{equation}
Hence, in quantum instrument CMM scaling of the update map $T_A(x)$ by probability $P_\rho(A=x)$ is a linear map. Generally, LSR-CMM with the context (state) space ${\cal C}= {\cal D}$   can be endowed with  context (state) update maps such that scaling  (\ref{SC}) can be nonlinear. CMM $M_{\rm{QI; NL}}$ with nonlinear context update maps might be useful for quantum-like modeling. It is interesting to find concrete applications of   $M_{\rm{QI; NL}}.$ Of course, such applications would lead to modification of the indirect measurement scheme.

 We can summarize properties of quantum instrument CMM, both models $M_{\rm{QI; DL}}$ and $M_{\rm{QI;O}}$:
\begin{itemize}
\item violation of FTP - yes
\item  OE - yes
\item  violation of replicability - yes
\item  RRE - no
\item   OE+RRE - generally no, but can be realized by special instruments 
\item violation of Bell inequalities - yes
\end{itemize}

It is interesting to find a property distinguishing  $M_{\rm{QI; DL}}$ and $M_{\rm{QI;O}}$ via an experimental test, i.e., some experimentally testable property such that only completely positive instruments have it.  

One of the important features of the von Neumann model  is coincidence of JPD and conditional JPD for compatible observables. In contrast,  the instrument model shows that generally the situation is not simple at all. Consider two instruments ${\cal I}_A(x)$ and  ${\cal I}_B(y)$ such that their observables are POVMs of PVM type, i.e., $\hat A=(\hat E_A(x))$ and $\hat B=(\hat E_B(y)).$ They are jointly measurable and the JPD is given by formula (\ref{JPD1}). The conditional JPD is given by
\begin{equation}
\label{JPD3}
P_\rho(A=x, B=y) = P_\rho(A=x) P_\rho(B=y|A=x)= 
\end{equation}  
$$
P_\rho(A=x) P_{\rho_{A=x}}(B=y)= \rm{Tr} {\cal I}_B(x) {\cal I}_A(x) \rho.   
$$
The right hand sides of (\ref{JPD1}) and (\ref{JPD3}) coincide only if the instrument superoperators are of the 
projection type, i.e.,  ${\cal I}(x)\rho= \hat E(x) \rho \hat E(x).$

Moreover, in this case two projection type observables, PVMs,  can have a variety of conditional probability distributions corresponding different instruments generating them by the rule 
$\hat E(x) = {\cal I}^\star(x).$

\section{Ordered Space Measurement Model with Probability Measure States}
\label{CLPR}

In this section we connect the generalized probability theory (the Davies-Lewis approach \cite{Davies-Lewis})  for probability measures with CMM. 
Here we use the ordered linear space approach. This is the concrete application of the universal  scheme based on the abstract framework of ordered linear spaces.  

Consider the space ${\cal M}$ of all real valued measures on some set $\Omega$ with a $\sigma$-algebra of subsets ${\cal F},$ i.e., ${\cal M}\equiv {\cal M} (\Omega,{\cal F}).$  Real linear space ${\cal M}$ has the natural order structure and the positive cone ${\cal M}^+$ consisting of non-negative measures. Consider the elements of this cone given by probability measures, i.e., $\mu\geq 0$ and $\mu(\Omega)=1;$ denote this set 
by the symbol ${\cal S};$ this is the set of states; ${\cal S}$ is a convex subset of ${\cal M}.$
The latter is endowed with the variation norm, $||\mu||= \rm{var}(\mu)$ and it is a Banach space. Consider its dual space 
${\cal M}^\prime,$ the space of continuous linear functionals $f: {\cal M} \to \mathbb{R}.$  Denote by 
${\cal A}$ the subset of  ${\cal M}^\prime,$ consisting of functionals mapping ${\cal S}$ into $[0,1].$ Elements of ${\cal A}$ are called effects, these are basic observables. They can be described  solely in the terms of the state space ${\cal S}$ as affine functionals  valued in $[0,1],$ i.e.,
${\cal A}\equiv {\cal A}({\cal S}).$

Consider the functional $u \in {\cal M}^\prime $  defined as $ \mu \to \langle u|\mu\rangle= \mu (\Omega).$ Its characteristic property is that $\langle u|\mu\rangle= 1$ for any state $\mu \in {\cal S}.$  

Let $X=\{x_1,..,x_m\}$ be a finite set and let $A=(A(x_i), i=1,...,m)$ where $A(x_i) \in {\cal A}({\cal S})$ and $A(X) \equiv \sum_{x \in X}  A(x) =u.$ Such vectors of functionals are analogs of POVMs; we call them ${\cal M}$-POVMs. These are observables of the contextual probability space $\Sigma_{\rm{measure}}$ with contexts ${\cal C}= {\cal S}$ and the set of probability distributions ${\cal P}$ defined as 
$$
P_\mu^A(x)\equiv P_C(A=x)= \langle A(x)| \mu\rangle.
$$ 

As we learn from the quantum instrument theory, the basic elements of measurement procedures are not observables, but instruments. Let ${\cal L}({\cal M})$ denote the space of continuous linear operators, $J: {\cal M} \to {\cal M}.$ The ${\cal M}$-instrument with the range of values $X$ is a map  ${\cal I}: X \to {\cal L}({\cal M})$ such that  ${\cal I}(x) {\cal M}^+ \subset {\cal M}^+$ and 
$$
{\cal I}(X)\equiv \sum_x {\cal I}(x) : {\cal S} \to {\cal S}.
$$          
Each instrument determines the state update map
$$
\mu \to T_A(x) \mu \equiv \frac{1}{{\cal I}(x) \mu(\Omega)}{\cal I}(x) \mu=
\frac{1}{\langle u|{\cal I}(x) \mu\rangle}{\cal I}(x) \mu,
$$
and the probability distribution 
$$
P_\mu(A=x)= \langle u|{\cal I}(x) \mu\rangle
$$
The domain of definition of the state update map $T_A(x)$ is given by the set of probability measures ${\cal C}_A(x)=\{ \mu \in S: P_\mu(A=x) >0 \}.$  

Let $J: {\cal M} \to {\cal M}$ be  a continuous linear operator. Then its adjoint operator $J^\star$ is well defined,  $J^\star: {\cal M}^\prime \to {\cal M}^\prime,$ and
$$
\langle J^\star f | \mu\rangle = \langle f  | J \mu\rangle.
$$
Set
$$
A(x)= {\cal I}^\star(x)  u.
$$
Then, for each $x,   A(x)$ is an effect, i.e. $A(x) \in {\cal A}( {\cal S}).$ 
So, each ${\cal M}$-instrument determines a ${\cal M}$-POVM. 

The ${\cal M}$-CMM consists of context (states) given by probability measures and POVM-observables
with state updates given by instruments.

\section{Linear Space Representation for Contextual Probability Space}
\label{MackeyS}

The state space is given by the set $S,$ the set of possible measurement outcomes 
of an observable quantity is denoted by $X.$ Let a
system be in a state $s \in S.$ A probability $p(x, s)$ is assigned to any possible outcome 
$x \in X.$ Thus, we have a function 
$$
p: X\times S \to [0,1].
$$
To each outcome $x \in X$ and state $s \in S,$ this function is a probability of the outcome $x$
for the system is in the state $s.$ Generalized probability model is a triple $(S,
 p(\cdot,\cdot), X).$  Denote by $\Phi_{[0,1]}$ the space of function from $X$ to $[0,1].$
By considering state $s$ as a variable, we obtain the map
\begin{equation}
\label{S1}
S \to \Phi_{[0,1]}, \; s\to s(\cdot)=p(\cdot,s).
\end{equation}
It is natural to assume that each state $s\in S$ determines the probability distribution uniquely, i.e., 
\begin{equation}
\label{L2m2}
p(x,s_1)= p(x,s_2)\; \mbox{ for any }   x \in X \Rightarrow s_1=s_2. 
\end{equation}
Under this assumption the map (\ref{S1}) is injection.
Thus, each state $s$ can be mapped to a function belonging to space $\Phi_{[0,1]},$ it will be denoted by the same symbol $s.$ Consider now the vector space $\Phi$ of all real valued functions on $X.$ So, $S$ is identified  with a subset of this functional space.
Consider its closed convex hull $\bar{S}.$ The vectors from it are all possible probabilistic mixtures (convex combinations) of states in $S.$  

Each $x \in X$ defines a linear functional on $\Phi, \;\phi \to f_x(\phi) = \phi(x).$ If $\phi=s \in \bar{S},$ 
then $f_x(s)=s(x) \in [0,1], $ i.e., 
$
f_x: \bar{S} \to [0,1].
$ 
This is an affine functional on the convex set $\bar{S}.$ It describes a measurement outcome and  $f_x(s)=p(x,s)$ is the probability for this outcome in state $s.$ 

Denote by $A(\bar{S})$ to the space of all affine functionals 
$$
f: \bar{S} \to [0,1].
$$
In particular, for any $x \in X, \; f_x \in  A(\bar{S}).$ 
Any functional $f \in A(\bar{S})$ describes an outcomes of some  observable, and thus $f (s)$ is the probability for that outcome in state $s.$ 

In QM, $\bar{S}$ is the set of density operators and elements of $A(\bar{S})$  are effects - components of POVMs, $ f(\rho)= \rm{Tr} \hat E_f  \hat \rho, $ where $\hat E_f$ is the effect corresponding to the affine functional $f.$  

The elements of $A(\bar{S})$ are called {\it effects}.  It is typically assumed that there exists an element $u$ of $A(\bar{S})$ such that 
$u(s)=1$ for any $s \in \bar{S}.$ It is an analog of quantum observable given by the unit operator $I.$
Consider the point wise order structure on $A(\bar{S}), f \leq g$ iff $f(s) \leq g(s)$ for any state $s.$
Thus, any observable $f \in A(\bar{S})$ is majorated by $u, 0 \leq f \leq u.$ A discrete measurement is
represented by a set of effects $(f_i)$ such that $\sum_i f_i = u.$

We now connect LSR for the contextual probabilistic model.  We assume that all observables have the same range of values $X.$ The straightforward intention is to set $S= {\cal C} \times {\cal O}$.  Let, as above, 
$\Phi_{[0,1]}$ denotes the space of functions from $X$ to $[0,1].$ We map $S$ into  $\Phi_{[0,1]}, \;
s=(C,A) \to P_C^A.$  However, generally this map is not injection: $P_{C_1}^{A_1}(x)= P_{C_2}^{A_2}(x)$  for all $x \in X$ does not imply that $C_1=C_2$ and $A_1=A_2.$ So, such straightforward construction seems to be non-proper for our aim. 

We modify it by setting ${\bf X}= {\cal O} \times X,$ the elements of ${\bf X}$ are pairs ${\bf x}=$(observable, outcome)$=(A, x).$ We now use the symbols $\Phi_{[0,1]}$ and $\Phi$ for functions from 
${\bf X} \to [0,1]$ and to real line respectively.  Each context $C$ can be represented as a vector belonging to $\Phi_{[0,1]},\; C({\bf x})=
C(A,x) = P_C^A(x).$  Due to (\ref{L2m1}), embedding of the set of contexts ${\cal C}$ into $\Phi_{[0,1]}$ is injection. Again denote 
by $\bar{\cal C}$ the convex hull of ${\cal C}.$  Each point  ${\bf x}=$(observable, outcome)$=(A, x)$ determines the affine functional $C \to f_{A,x}(C)= P_C^A(x) \in [0,1].$ 
Now fix $A\in {\cal O}$ and consider the family of functionals $F= (F_A(x)= f_{A,x}: x \in X).$ 
This is representation of observable $A.$ 

So, any contextual probability model can be realized alike COM -- an observational COM.

\section{Concluding remarks}

As was emphasized in introduction, CMM can be considered as the most general probabilistic model for measurement. It also can be considered as the minimalist restructuring of Mackey's project \cite{Mackey}. Mackey proceeded to quantum logic and this makes the mathematical construction more complicated. One may even say that mathematics shadowed measurement theory. Surprisingly,  even this minimalist model (CMM) has a complex structure and  represents the basic elements of quantum probability and measurement theory,e.g., interference of probability, order effect, entanglement, the violation of the Bell inequalities.

CMM can be employed not only in quantum foundations, but also in quantum-like modeling that  can employ  contextual probability calculi and CMMs which are not based on the complex Hilbert space formalism.

\section*{Acknowledgments}

This paper is the completion of the long project on contextual probability and quantum physics 
which started with my discussions with Kolmogorov and Mackey and later with 
Accardi, Ballentine, Gudder, Mittelstaedt, Ohya, Shiryaev, Volovich. During the recent years, I discussed with Plotnitsky the Bohr's 
complementarity principle  and with Ozawa quantum instrument realization of measurement theory. Since year 2000, I was involved in critical and stimulated debates with Fuchs on QBism and subjective probability in QM. During my visits to Vienna, I had exciting conversations with Rauch and Zeilinger on the (non)realism, (non)locality,  and (non)contextuality of QM.
All these discussions stimulated my thinking on contextual measurement theory and probability.

\section*{Apepndix 1. Terminology: Context vs. State}

We make the following remark about the terminology ``context vs. state.'' Since QM operates with the notion ``state'', generalized  probability theory also employes this terminology. However, even in QM  using the term ``state'' is ambiguous. It matches the orthodox Copenhagen interpretation by which a state is treated as the state of an individual quantum system, say the state of an electron - one concrete electron. Many experts consider this interpretation of the quantum state as leading to paradoxes and mismatching with the statistical nature of quantum phenomena. This is a complicated foundational issue, since the leading supporters of the orthodox Copenhagen interpretation also consider QM as a statistical theory in that the state of an individual system encodes the statistics of the coming experimental runs.
  
	For example, 
Einstein, Koopman, Margenau, Blohintzev, Ballentine, and nowadays, e.g.,  Ballian, Nieuwenhuizen, Khrennikov use the so called statistical (or ensemble) interpretation of QM. By this interpretation a quantum state represents statistical properties of an ensemble of identically prepared systems. So, whose state? The state of an ensemble? In the operational approach ``state'' corresponds to a preparation procedure. It seems that the term ``state'' borrowed from the orthodox Copenhagen interpretation does not match to the statistical and operational interpretations of QM. In the generalized probability theory the term ``state'' is typically 
associated with a preparation procedure or a class of equivalent preparation procedures. However, this meaning of the state is not highlighted and the output of the generalized probability theory is often projected onto the the orthodox Copenhagen interpretation, i.e., this theory interpreted as a theory about the structure of the state space of individual quantum systems. Therefore in the V\"axj\"o interpretation we prefer to use the notion of context as a complex of experimental conditions, pre-measurement context can be associated with a class of equivalent preparation procedures (as is done in the consistent presentation of the generalized probability theory), measurement context is the combination of the preparation, measurement, and state update generated by measurement feedback with the fixed outcome.      

In contrast to the generalized probability theory employing LSR, we do not assume that the set of pre-measurement  contexts ${\cal C}$ contains contexts generated by statistical mixtures (see Axiom 4 in Mackey's book \cite{Mackey}), i.e., for $C_1, C_2 \in {\cal C}$ and $p_1, p_2 \geq 0, p_1+p_2=1,$ the set ${\cal C}$ need not contain a context which can be identified with $p_1 C_1+p_2 C_2.$  Proceeding without the mixture axiom illuminates the difference between state and context; consider e.g., the ``basic contextual probability representation'' of the classical Kolmogorov probability space (section \ref{KKK}). Here contexts are not probability distributions, but elements of the ($\sigma$-)algebra. Generally a context provides a finer description of the measurement setup than a probability distribution. 

\section*{Apepndix 2. Contextual measurement model with vs. without linear space representation}

{\it Why is it useful to proceed in contextual probabilistic framework as far as possible without appealing to LSR? }

I start with some remarks on uncritical using of LSR:
\begin{itemize}
\item LSR shadows the essence of the quantum probability formalism as the machinery for probability inference.
\item LSR for classical probability via the use of the linear space of measures with the positive cone of non-negative measures and convex state space of probability measures seems to be inadequate to 
Kolmogorov's  theory \cite{K,KE} based on conditioning (contextualization) with Bayes' formula (section \ref{KKK}).  
\item LSR generates (via creation of convex linear hull and its closure) a plenty of unphysical states and observables (see Ballentine \cite{BL}), operating with them led e.g. to von Neumann's no-go theorem \cite{VN}.\footnote{ Generally some outputs of quantum information theory obtained in the  abstract LSR framework might be its artifacts, without coupling to physical reality. The critical analysis of connection of LSR mathematics and physics is needed.} 
\item The picture that quantum probability theory is just one of LSR of probability diminishes the exclusiveness of linearity of QM. One losses the physical ground for the latter, LSR becomes just a part of
the mathematical apparatus of QM (see section \ref{QB} on my views on the physical ground for QM-linearity).
\item Linking of entanglement to the LSR tensor product structure shadows its contextual probabilistic nature and supports the ambiguous statements on quantum nonlocality.  
\item Recently the mathematical formalism of quantum theory, especially probability, started to be 
widely applied outside of physics, e.g., in cognition, psychology, social and political sciences, economics and finance, so called quantum-like modeling (see, e.g., recent monograph \cite{Open_KHR}). In such models the set of possible states 
(pre-measurement contexts) is not so rich as in physics. In quantum-like modeling even the possibility to prepare 
statistical mixtures is not evident, i.e., proceeding towards convex structures might be misleading.
\end{itemize}


\begin{thebibliography}{999}

\bibitem{VN0} Von Neumann, J.  {\it  Mathematische Grundlagen der Quantenmechanik.} Springer-Verlag, Berlin, 1932.
\bibitem{FeynmanP} Feynman, R. P.  (1951). \emph{The Concept of probability in quantum mechanics}; Berkeley Symp. on Math. Statist. and Prob. Proc. Second Berkeley Symp. on Math. Statist. and Prob. (Univ. of Calif. Press), 533--541.
\bibitem{Feynman} Feynman, R. and Hibbs, A.: Quantum Mechanics and Path Integrals. McGraw-Hill, New York (1965)
\bibitem{Koopman} Koopman, B.O.  1955. Quantum theory and the foundations of probability. In Applied Probability. L. A. MacColl, Ed.: 97-102. McGraw-Hill. New York.

\bibitem{Mackey0} Mackey, G.W. Quantum mechanics and Hilbert space. Am. Math. Mon. 1957, 64, 45–57.
\bibitem{Mackey}  Mackey, G. N. (1963).  \emph{Mathematical foundations of quantum mechanics.} Benjamin, INC, New York, Amsterdam.
 
\bibitem{Davies-Lewis} Davies, E. B. and Lewis, J. T. An operational approach to quantum probability. Commun. Math. Phys. 17, 239–260 (1970).
\bibitem{DV}   Davies, E. B.  {\it Quantum theory of open systems},  (Academic Press,  London, 1976).

\bibitem{Gudder} Gudder, S.P. Convex structures and operational quantum mechanics. Commun. Math. Phys. 29, 249–264 (1973).
\bibitem{Gudder1} Gudder, S.P. Stochastic Methods in Quantum Mechanics; Courier Corporation: Mineola, NY, USA, 2014.

\bibitem{Ozawa} Ozawa, M. Optimal measurements for general quantum systems. Rep. Math. Phys. 18, 11–28 (1980).
\bibitem{Oz1} Ozawa, M.  Probabilistic interpretation of quantum theory. {\it New Generation Computing} {\bf 34}, 125-152 (2016).

\bibitem{Accardi1} Accardi, L. Accardi, L. (1981). Topics in quantum probability. Phys. Rep., 77(3), 169-192.
\bibitem{Accardi2} Accardi, L. (1984). The probabilistic roots of the quantum mechanical paradoxes. In The Wave-Particle Dualism: A Tribute to Louis de Broglie on his 90th Birthday (pp. 297-330). Dordrecht: Springer Netherlands.
\bibitem{Accardi3} Accardi, L.  Urne e camaleonti. Il Saggiatore, Rome, 1997.
\bibitem{Accardi4} Accardi, L. (2022). New Challenges for Classical and Quantum Probability. Entropy, 24(10), 1502.

\bibitem{BL0} L. Ballentine, Probability in Quantum Mechanics. Annals of New York Academy of Science, Techniques and Ideas in
 Quantum Measurement Theory, 480, N 1, 382-392 (1986).
\bibitem{BL} Ballentine, L. E. (1989).  The statistical interpretation of quantum mechanics, \emph{Rev. Mod. Phys.},  42, 358--381.
\bibitem{BL2} Ballentine, L. E. (2014). Quantum mechanics: a modern development. WSP, Singapore.
\bibitem{BL2a} Ballentine, L. E.:  Interpretations of probability and quantum theory. In:  Khrennikov, A. Yu. (ed)  Foundations of
Probability and Physics, {\it Quantum Probability and White Noise Analysis} {\it  13},  71-84. WSP: Singapore, 2001.

\bibitem{Svozil} Svozil, K. Quantum Logic; Springer Science and Business Media: New York, NY, USA, 1998.

\bibitem{INT} Khrennikov,  A.~Yu.  \emph{Interpretations of Probability}; VSP Int. Sc. Publishers: Utrecht/Tokyo, 1999;
 2nd edn., De Gruyter: Berlin, 2009.

\bibitem{Knuth1} Goyal, P.; Knuth, K.H.; Skilling, J. Origin of complex quantum amplitudes and Feynman’s rules. Phys. Rev. A 2010, 81, 022109.
\bibitem{Holik} Holik, F., Massri, C., Plastino, A., and Saenz, M. (2021). Generalized probabilities in statistical theories. Quantum Reports, 3(3), 389-416.

\bibitem{KHR2} Khrennikov,  A.~Yu.: Origin of quantum probabilities. In: Khrennikov, A.(ed.)  Foundations of Probability and Physics, pp. 180--200.  V\"axj\"o-2000, Sweden; WSP, Singapore (2001).
\bibitem{KHR3a} Khrennikov, A.: Contextual viewpoint to quantum stochastics. J. Math. Phys. \textbf{44}(6), 2471-2478  (2003)
\bibitem{KHR3b} Khrennikov, A.: Representation of the Kolmogorov model having all distinguishing features of quantum probabilistic model. Phys. Lett. A \textbf{316}(5), 279-296  (2003)
\bibitem{KHR_interference} Khrennikov, A. Y. (2007). A formula of total probability with the interference term and the Hilbert space representation of the contextual Kolmogorovian model. Theor. Prob. and Appl. 51(3), 427-441.
\bibitem{KHR_CONT} Khrennikov,  A. (2009). \emph{Contextual Approach to Quantum Formalism}, (Springer, Berlin-Heidelberg-New York).

\bibitem{VM} von Mises, R. (1957). \textit{Probability, statistics and truth.} London: Macmillan.

\bibitem{Vaxjo2002}  Khrennikov, A.:  V\"axj\"o interpretation of quantum mechanics. In: Khrennikov, A. (ed.) Quantum Theory: Reconsideration of Foundations,  Ser. Math. Modelling  \textbf{2}, pp. 163--170.  V\"axj\"o Univ. Press, V\"axj\"o (2002);  arXiv:quant-ph/0202107
\bibitem{V2}  Khrennikov, A.  V\"axj\"o interpretation-2003: Realism of contexts.  In  \emph{Quantum Theory: Reconsideration of Foundations}; Khrennikov A., Ed.; V\"axj\"o Univ. Press: V\"axj\"o, 2004, pp. 323--338.

\bibitem{HAVKHRQ}  Haven, E. and Khrennikov, A.:  Quantum probability and the mathematical modelling of decision-making. Phil. Trans. Royal Soc. A \textbf{374}(2058), 20150105 (2016)
\bibitem{HAVKHRQ1} Haven, E., and Khrennikov, A.:  Statistical and subjective interpretations of probability in quantum-like models of cognition and decision making. J. Math. Psych. 74, 82-91 (2016)

\bibitem{K} A. N. Kolmogoroff. Grundbegriffe der Wahrscheinlichkeitsrechnung. Springer, Berlin (1933).
\bibitem{KE} Kolmogorov, A. N.: Foundations of the Theory of Probability.  Chelsea Publ. Company, New York (1956).
\bibitem{VN} Von Neumann, J. {\it Mathematical Foundations of Quantum Mechanics}; Princeton University Press: Princeton, NJ,
USA, 1955.

\bibitem{QL1} Khrennikov, A.: Ubiquitous Quantum Structure: from Psychology to Finances. Springer, Berlin-Heidelberg-New York  (2010).
\bibitem{KHRfrontiers} Khrennikov A. Quantum-like model of unconscious-conscious dynamics. Front. Psych. \textbf{6},  997--1010 (2015).
\bibitem{Open_KHR} Khrennikov, A. Y. (2023). Open Quantum Systems in Biology, Cognitive and Social Sciences. Springer Nature.

\bibitem{BR0} Bohr, N.    \emph{The Philosophical Writings of Niels Bohr};  Ox Bow Press: Woodbridge, UK, 1987.

\bibitem{NL0B}  Khrennikov, A., Bohr against Bell: complementarity versus nonlocality. {\it Open Phys.} {\bf 2017}, {\it 15}, 734-773.
\bibitem{ABELL} Khrennikov, A. After Bell. {\it Fortschritte der Physik (Progress in Physics)}.
{\bf 2017},  {\it 65},  N 6-8, 1600014.

\bibitem{Bell2} Bell,  J.S.  On the problem of hidden variables in quantum theory.  {\it Rev. Mod. Phys.} {\bf 1966}, {\it 38}, 450.
\bibitem{Bell1} Bell, J.S. \textit{Speakable and Unspeakable in Quantum Mechanics}, 2nd ed.;  Cambridge University Press: Cambridge, UK, 2004.  

\bibitem{Beltrametti} Beltrametti, E.G. and Cassinelli, C.: The logic of quantum mechanics. SIAM  \textbf{25}, 429--431 (1983)

\bibitem{O1} Ozawa, M.:   Quantum measuring processes for continuous observables. J. Math. Phys.  \textbf{25}, 79--87 (1984)
\bibitem{O6} M. Ozawa, On information gain by quantum measurements of continuous observables, J. Math. Phys. 27, 759-763 (1986).
\bibitem{O7} M. Ozawa, Mathematical characterizations of measurement statistics, Quantum Communication and Measurement, Plenum, 109-117 (1995). 
\bibitem{O3} Ozawa, M.:   An operational approach to quantum state reduction. Ann. Phys.  (N.Y.) \textbf{259}, 121--137 (1997)

\bibitem{dariano1} Chiribella, G., D'Ariano, G. M., and Perinotti, P. (2009). Realization schemes for quantum instruments in finite dimensions. Journal of mathematical physics, 50(4).
\bibitem{dariano2} D'Ariano, G., M., Chiribella, G., Perinotti, P.:   Quantum Theory from First Principles: An Informational Approach. Cambridge Univ. Press, Cambridge (2017)
\bibitem{dariano3} D'Ariano, G. M., Perinotti, P., and Tosini, A. (2022). Incompatibility of observables, channels and instruments in information theories. J. Phys. A: Math. Theor., 55(39), 394006.

\bibitem{KHR_iterference} Khrennikov, A. Y. (2007). A formula of total probability with the interference term and the Hilbert space representation of the contextual Kolmogorovian model. Theor. Prob. and Appl. 51(3), 427-441.
\bibitem{LF} Khrennikov, A. Y., and Loubenets, E. R. (2004). On relations between probabilities under quantum and classical measurements. Foundations of Physics, 34, 689-704.

\bibitem{Fuchs1} Fuchs, C.~A.:  Quantum mechanics as quantum information (and only a little more). In: Khrennikov, A.  (ed.),  Quantum Theory: Reconsideration of Foundations, Ser. Math. Model. \textbf{2}, pp. 463--543. V\"axj\"o Univ. Press, V\"axj\"o (2002) 
\bibitem{Fuchs2} Fuchs, C.~A.: The anti-V\"axj\"o interpretation of quantum mechanics. In: Khrennikov, A.  (ed.), Quantum Theory: Reconsideration of Foundations, Ser. Math. Model. \textbf{2}, pp. 99--116. V\"axj\"o Univ. Press, V\"axj\"o (2002)
\bibitem{Fuchs3} Fuchs, C. A. and Schack, R.: A quantum-Bayesian route to quantum-state space. Found. Phys. \textbf{41}, 345–356 (2011)
\bibitem{Fuchs5} Fuchs, C. A. (2023). QBism, Where Next?. arXiv preprint arXiv:2303.01446.


\bibitem{WB} Wang, Z. and Busemeyer, J.~R.:  A quantum question order model supported by empirical
tests of an a priori and precise prediction.  Top. Cogn. Sc. \textbf{5}, 689--710 (2013).
\bibitem{WSSB14} Wang, Z., Solloway, T., Shiffrin, R.~M. , and Busemeyer, J.~R.:  Context effects produced by question orders reveal quantum nature of human judgments. PNAS \textbf{111}, 9431--9436 (2014). 
\bibitem{PLOS} Khrennikov, A.,  Basieva, I., Dzhafarov, E.N., and   Busemeyer, J. R.: Quantum models for psychological measurements: An unsolved problem. PLOS ONE \textbf{9}, Art. e110909 (2014).
\bibitem{ENTROPY} Ozawa, M. and  Khrennikov, A.:   Application of theory of quantum instruments to psychology: Combination of question order effect with response replicability effect. Entropy, \textbf{22}(1), 37 (2020) 1-9436.
\bibitem{OJMP} Ozawa, M. and Khrennikov, A.: Modeling combination of question order effect, response replicability effect, and QQ-equality with quantum instruments. J. Math. Psych. \textbf{100}, 102491  (2021).



\bibitem{ENT} Basieva, I., and Khrennikov, A. (2022). Conditional probability framework for entanglement and its decoupling from tensor product structure. J. Phys. A: Math. Theor., 55(39), 395302.
\bibitem{ENT1} Basieva, I., and Khrennikov, A. Entanglement of observables: Quantum conditional probability approach. Found. Phys., 53, Article number: 84 (2023).

\bibitem{EPR} Einstein, A.; Podolsky, B.; Rosen, N.   Can quantum-mechanical description of physical reality be considered complete? \emph{Phys. Rev.} {\bf 1935}, {\it 47}, 777--780.

\bibitem{SCHE} Schr\"odinger, E.: Die gegenw\"artige Situation in der Quantenmechanik, Naturwissenschaften \textbf{23}, 807–-812, 823–-828, 844–849 (1935).
\bibitem{SCHEa} Schr\"odinger, E.: The present situation in quantum mechanics: A translation of Schr\"odinger's ``Cat Paradox'' paper (by: J. D. Trimmer). Proc. Am. Philos. Soc. \textbf{124}, 323–-338 (1980).


\bibitem{KHRWS} Khrennikov, A.:  Probability and Randomness. Quantum versus Classical. WSP, Singapore (2016)


\end{thebibliography}
\end{document}